\documentclass[11pt]{article}
\DeclareMathAlphabet{\scr}{U}{rsfs}{m}{n}

\usepackage{latexsym}
\usepackage{epsfig}
\usepackage[mathscr]{eucal}
\usepackage{amsfonts}
\usepackage{amscd}
\usepackage{cite}
\usepackage{array}   
\usepackage{amssymb}
\usepackage{colordvi}
\usepackage[centertags]{amsmath}
\usepackage{enumerate}
\usepackage{graphicx}
\usepackage{booktabs}
\usepackage{theorem}
\usepackage[footnotesize]{caption}
\usepackage{soul}
\usepackage{mcite}
\usepackage{slashed}
\usepackage{color}
\usepackage{ulem}
\newcommand{\newc}{\newcommand}
\newc{\be}{\begin{equation}}
\newc{\ee}{\end{equation}}
\newc{\bea}{\begin{eqnarray}}
\newc{\eea}{\end{eqnarray}}
\newc{\ol}{\overline}
\newc{\wt}{\widetilde}
\newc{\bs}{\boldsymbol}
\newc{\m}{\mathcal}
\newc{\la}{\langle}
\newc{\ra}{\rangle}

\newcommand{\non}{\nonumber} 
\newcommand{\beq}{\begin{eqnarray}} 
\newcommand{\eeq}{\end{eqnarray}} 
\newcommand{\bpmatrix}{\begin{pmatrix}}
\newcommand{\epmatrix}{\end{pmatrix}}
\newcommand{\ba}{\begin{array}}
\newcommand{\ea}{\end{array}}
\newcommand{\braket}[1]{\left(#1\right)}

\newcommand{\fr}{\frac}
\newcommand{\hc}{\text{ h.c.}}
\newcommand{\diag}{\text{diag}}
\newcommand{\al}{\alpha}

\newcommand{\MH}{\mathcal{M}_{hh}}

\newcommand{\drbar}{\overline{\text{DR}}}
\renewcommand{\ol}{\text{1l}}

\renewcommand{\Re}{\text{Re}\!}

\newcommand{\order}{\mathcal{O}(\alpha_t\alpha_s)}
\newcommand{\deltatwo}{\delta^{ \text{\tiny(}\!\text{\tiny2}\!\text{\tiny)}}\!}
\newcommand{\deltaone}{\delta^{ \text{\tiny(}\!\text{\tiny1}\!\text{\tiny)}}\!}
\newcommand{\wavetwo}{\deltatwo\mathcal{Z}}
\newcommand{\mueff}{\mu_{\text{eff}}}
\newcommand{\figref}[1]{Fig.~\ref{#1}}
\renewcommand{\eqref}[1]{Eq.~(\ref{#1})}
\newcommand{\bib}[1]{Ref.~\cite{#1}}

\newcommand{\ssect}[1]{Subsection~\ref{#1}}

\newcommand{\appen}[1]{Appendix~\ref{#1}}

\newcommand{\DRb}{\overline{\text{DR}}}

\newcommand{\OS}{\text{OS}}
\newcommand{\MSb}{\overline{\text{MS}}}

\newcommand{\ie}{{\it i.e.\;}}
\newcommand{\eg}{{\it e.g.\;}}
\newcommand{\bc}{\begin{center}}
\newcommand{\ec}{\end{center}}
\newcommand{\hs}{\hspace*{3mm}}
\newcommand{\gev}{~\text{GeV}}
\newcommand{\mev}{~\text{MeV}}

\newcommand{\s}{\newline \vspace*{-3.5mm}}

\begin{document}

\title{
\vspace*{-3cm}
\phantom{h} \hfill\mbox{\small FR-PHENO-2014-012}\\[-1.1cm]
\phantom{h} \hfill\mbox{\small KA-TP-32-2014}\\[-1.1cm]
\phantom{h} \hfill\mbox{\small SFB/CPP-14-95} 
\\[1cm]
\textbf{Two-Loop Contributions of the Order $\order$\\
 to the Masses of the Higgs Bosons \\ in the CP-Violating NMSSM \\[4mm]}}

\date{}
\author{
Margarete M\"{u}hlleitner$^{1\,}$\footnote{E-mail:
  \texttt{margarete.muehlleitner@kit.edu}} ,
Dao Thi Nhung$^{2\,}$\footnote{E-mail: \texttt{thi.dao@kit.edu}} ,
Heidi Rzehak$^{3\,}$\footnote{E-mail:
  \texttt{heidi.rzehak@physik.uni-freiburg.de}} ,
Kathrin Walz$^{1\,}$\footnote{E-mail: \texttt{kathrin.walz@kit.edu}}
\\[9mm]
{\small\it
$^1$Institute for Theoretical Physics, Karlsruhe Institute of Technology,} \\
{\small\it 76128 Karlsruhe, Germany.}\\[3mm]
{\small\it
$^2$Institute of Physics, Vietnam Academy of Science and Technology,} \\
{\small \it 10 DaoTan, BaDinh, Hanoi, Vietnam.}\\[3mm]
{\small\it
$^3$ Physikalisches Institut, Albert-Ludwigs-Universit\"{a}t Freiburg,}\\
{\small\it 79104 Freiburg, Germany.}\\[3mm]
}

\maketitle

\begin{abstract}
\noindent
We provide the two-loop corrections to the Higgs boson masses of the
CP-violating NMSSM in the Feynman diagrammatic approach with vanishing
external momentum at ${\cal O} (\al_t \al_s)$. The adopted
renormalization scheme is a mixture between $\DRb$ and on-shell
conditions. Additionally, the renormalization of the top/stop sector
is provided both for the $\DRb$ and the on-shell scheme. The calculation is
performed in the gaugeless limit.  We find that the two-loop
corrections compared to the one-loop corrections are of the order
of 5-10\%, depending on the top/stop renormalization scheme. The
theoretical error on the Higgs boson masses is reduced due to the
inclusion of these higher order corrections. 
\end{abstract}
\thispagestyle{empty}
\vfill
\newpage
\setcounter{page}{1}

\section{Introduction}
The discovery of the Higgs boson by the LHC experiments ATLAS
\cite{Aad:2012tfa} and CMS \cite{Chatrchyan:2012ufa}
has been a milestone in our quest for understanding the origin of
particle masses. While the investigation of the properties of this
scalar particle strongly suggests that it is the Higgs boson of the
Standard Model (SM), the present precision of the experimental data
still leaves room for interpretations in extensions beyond the SM
(BSM). Among these, models based on supersymmetry (SUSY) certainly
rank among the most intensely studied SM extensions. Supersymmetry allows
to cure some of the flaws of the SM. Thus 
{\it e.g.} the symmetry between bosonic and fermionic degrees of freedom 
solves the hierarchy problem, the inclusion of $R$-parity leads to a
possible dark matter candidate and the possibility of
additional sources for CP violation provides one of the three
necessary conditions for successful baryogenesis. Up to now, however,
no SUSY particles have been discovered, and the LHC has put lower
limits of around 1.5~TeV on the gluino mass and the squark
masses of the first two generations. On the other hand, from analysis
strategies based on monojet-like and charm-tagged event selections it
can be concluded that the mass of the lightest stop can still be rather light 
\cite{CMSPAS13,Aad:2014kra,Aad:2014nra,Boehm:1999tr,Hikasa:1987db,Muhlleitner:2011ww,Grober:2014aha},
down to about 240~GeV for arbitrary neutralino masses \cite{Aad:2014nra}. The 
stops provide the dominant contribution to the Higgs mass corrections
and play a crucial role in pushing the mass of the SM-like SUSY Higgs
boson to the necessary 126~GeV. In the minimal
supersymmetric extension (MSSM)\cite{Gunion:1989we,Martin:1997ns,Dawson:1997tz,Djouadi:2005gj} this requires large values of the stop
masses and/or mixing and thus challenges the naturalness of the model
due to fine-tuning. The situation is relaxed in the next-to-minimal SUSY
extension (NMSSM) \cite{Fayet:1974pd,Barbieri:1982eh,Dine:1981rt,Nilles:1982dy,Frere:1983ag,Derendinger:1983bz,Ellis:1988er,Drees:1988fc,Ellwanger:1993xa,Ellwanger:1995ru,Ellwanger:1996gw,Elliott:1994ht,King:1995vk,Franke:1995tc,Maniatis:2009re,Ellwanger:2009dp}: new contributions to the quartic coupling
stemming from the introduction of a complex superfield, which couples
with the strength $\lambda$ to the two Higgs doublet superfields
present in the MSSM, shift the tree-level mass of the lightest CP-even 
MSSM-like Higgs boson to a higher value. Therefore smaller loop
corrections are required to attain the measured Higgs mass value, and
lighter stop masses can generate a Higgs spectrum in accordance with the
experimental data (see {\it e.g.}~\cite{King:2012is,King:2012tr}). \s

In addition the NMSSM has many other interesting features. It can
incorporate CP violation in the Higgs sector already at tree level. The Higgs
spectrum may contain Higgs masses that are lighter than 126~GeV
without being in conflict with the experimental data, and allowing
{\it e.g.}~for substantial Higgs-to-Higgs decay widths \cite{Ellwanger:2013ova,Munir:2013dya,King:2014xwa,Bomark:2014gya}. Also situations
with two degenerate Higgs bosons around 126~GeV are possible
\cite{King:2012is,King:2012tr,Gunion:2012gc}. This
small list already gives a flavour of the plethora of interesting
phenomena that are possible in non-minimal SUSY phenomenology. On
the other hand it also shows the necessity of precise predictions for 
the Higgs mass and self-coupling parameters and for the production and the
decay processes, {\it i.e.}~including higher order calculations. In particular in the
Higgs sector there has been a lot of activity in pushing the
accuracy in the mass calculations to a level comparable to the one
achieved in the MSSM. In the CP-conserving NMSSM the leading one-loop
(s)top and (s)bottom contributions have been computed in
\cite{Ellwanger:1993hn,Elliott:1993ex,Elliott:1993uc,Elliott:1993bs,Pandita:1993tg}
and the chargino, neutralino as well as scalar one-loop contributions at
leading logarithmic accuracy have been provided by
\cite{Ellwanger:2005fh}. The full one-loop contributions in the $\DRb$
renormalization scheme have first been given in~\cite{Degrassi:2009yq}
and subsequently in~\cite{Staub:2010ty}. The 
authors of \cite{Degrassi:2009yq} have also provided the order ${\cal O}
(\alpha_t \alpha_s + \alpha_b \alpha_s)$ corrections in the approximation of zero
external momentum. Recently, first corrections beyond order ${\cal O}
(\alpha_t \alpha_s + \alpha_b \alpha_s)$ have been given in \cite{Goodsell:2014pla}.
We have furthermore calculated the full one-loop
corrections in the Feynman diagrammatic approach in a mixed $\DRb$-on-shell
and in a pure on-shell renormalization scheme \cite{Ender:2011qh}.  In the
mixed $\DRb$-on-shell renormalization scheme also the one-loop 
corrections to the Higgs self-couplings are available \cite{Nhung:2013lpa}.
CP-violating effects in the mass corrections have been considered in
Refs.~\cite{Ham:2001kf,Ham:2001wt,Ham:2003jf,Funakubo:2004ka,Ham:2007mt},
where contributions from the third generation squark sector, from the
charged particle loops and from gauge boson contributions have been 
computed in the effective potential approach at one loop-level. The
full one-loop and logarithmically enhanced two-loop effects have been
made available in the renormalization group approach
\cite{Cheung:2010ba}. We have complemented these calculations by
computing the full one-loop corrections in the Feynman diagrammatic
approach \cite{Graf:2012hh}. \s 

There are several codes available for the evaluation of the NMSSM mass
spectrum from a user-defined input at a user-defined scale. Thus
 {\tt NMSSMTools}
\cite{Ellwanger:2004xm,Ellwanger:2005dv,Ellwanger:2006rn} calculates
the masses and decay widths in the 
CP-conserving $\mathbb{Z}^3$. It can be interfaced with {\tt SOFTSUSY}
\cite{Allanach:2001kg,Allanach:2013kza}, which generates 
the mass spectrum for a CP-conserving 
NMSSM including the possibility of $\mathbb{Z}^3$ violation. The
interface of {\tt SARAH}
\cite{Staub:2010jh,Staub:2012pb,Staub:2013tta,Goodsell:2014bna,Goodsell:2014pla}
 with {\tt SPheno} \cite{Porod:2003um,Porod:2011nf} on the 
other hand allows for spectrum generations of different SUSY models,
including the NMSSM. In the same spirit, {\tt SARAH} has been
interfaced with the recently published package {\tt FlexibleSUSY}
\cite{Athron:2014yba,Athron:2014wta}. All these programs include the
Higgs mass corrections up to two-loop order, where in particular the
two-loop corrections are obtained in the effective potential approach. 
The program package {\tt NMSSMCALC}
\cite{Baglio:2013vya,Baglio:2013iia} for the calculation of the NMSSM
Higgs masses and decay widths, incorporates the one-loop corrections
in the full Feynman diagrammatic approach both for the CP-conserving
and CP-violating NMSSM. \s 

With the present work we contribute to the effort of achieving higher
precision in the computation of the NMSSM Higgs boson masses. We provide the
two-loop corrections to the neutral NMSSM Higgs boson masses in the
Feynman diagrammatic approach for zero 
external momentum at the order ${\cal O} (\al_t \al_s)$ based on a
mixed $\DRb$-on-shell renormalization scheme. In contrast
to the available results in the effective potential approach we 
calculate the two-loop corrections not only for the CP-conserving but
also for the CP-violating case. In the former case we find full agreement
with the results presented in~\cite{Degrassi:2009yq}. Our calculation
is performed in the gaugeless limit \ie we set the electric charge and
the $W$ and $Z$ boson masses to zero, $e=0, M_W=0, M_Z=0$. The vacuum
expectation value $v$ and the weak angle $\theta_W$
are kept at their SM values. Furthermore we neglect the bottom mass. 
These two-loop mass corrections have been included in the 
program package {\tt NMSSMCALC}. \s

The outline of our paper is as follows. 
In section \ref{sec:nmssmintro} we introduce the Higgs sector of the
CP-violating NMSSM, and we discuss in particular the quark and
squark sector, necessary for the order ${\cal O} (\al_t \al_s)$ corrections,
together with its renormalization. Section
\ref{sec:calculation} is dedicated to the calculation of the
mass corrections. Besides presenting the diagrams contributing to the
calculation, the counterterms and the applied renormalization
prescription are discussed in detail. We furthermore comment on the
tools we have used and the checks that we have performed to validate
our results. The numerical analysis is deferred to section
\ref{sec:numerical}. We show the impact of the two-loop corrections
along with the new features that appear with respect to the MSSM. An
estimate of the missing higher order corrections is given by applying 
two different renormalization schemes in the top (s)quark sector. We
summarize in section \ref{sec:conclusion}. 

\section{The CP-violating NMSSM \label{sec:nmssmintro}}
In order to set up our notation, we summarize here the main features 
of the complex NMSSM, concentrating on those parts of the Lagrangian,
that are relevant for the calculation of the  ${\cal
  O}(\al_t\al_s)$ corrections to the Higgs boson masses, {\it
  i.e.}~the Higgs and the stop sectors. For further details and
information on other sectors of the CP-violating NMSSM, see
\bib{Graf:2012hh}. We work in the framework of the NMSSM 
with a scale invariant superpotential and a discrete $\mathbb{Z}^3$ symmetry. In
terms of two Higgs doublet superfields $\hat{H}_d$ and $\hat{H}_u$, a
Higgs singlet superfield $\hat{S}$, the quark and lepton superfields
and their charged conjugates (denoted by the superscript $c$),
$\hat{Q}, \hat{U}^c, \hat{D}^c, \hat{L}, \hat{E}^c$, the NMSSM superpotential reads
\be  W_{NMSSM} = \epsilon_{ij} [y_e \hat{H}^i_d \hat{L}^j \hat{E}^c + y_d
\hat{H}_d^i \hat{Q}^j \hat{D}^c - y_u \hat{H}_u^i \hat{Q}^j \hat{U}^c] - \epsilon_{ij} \lambda \hat{S} \hat{H}^i_d
\hat{H}^j_u + \frac{1}{3} \kappa \hat{S}^3 \;. \ee 
The indices of the $SU(2)_L$ fundamental representation are denoted by
$i,j=1,2$, and $\epsilon_{ij}$ is the totally antisymmetric tensor
with $\epsilon_{12}= \epsilon^{12} = 1$. Here and in the following the
summation over equal indices is implicit. The colour and generation
indices have been suppressed. The dimensionless parameters $\lambda$
and $\kappa$ are considered to be complex in general. We throughout neglect
generation mixing, so that the Yukawa couplings $y_e, y_d, y_u$ are
diagonal and possible complex phases can be reabsorbed by redefining
the quark fields without changing the physical meaning \cite{Kobayashi:1973fv}. \s 

The soft SUSY breaking Lagrangian of the NMSSM expressed in terms of the scalar
component fields $H_u, H_d$ and $S$ reads
\begin{align}
{\cal L}_{\text{soft},\text{ NMSSM}} =& -m_{H_d}^2 H_d^\dagger H_d - m_{H_u}^2
H_u^\dagger H_u -
m_{\tilde{Q}}^2 \tilde{Q}^\dagger \tilde{Q} - m_{\tilde{L}}^2 \tilde{L}^\dagger \tilde{L}
- m_{\tilde{u}_R}^2 \tilde{u}_R^* 
\tilde{u}_R - m_{\tilde{d}_R}^2 \tilde{d}_R^* \tilde{d}_R 
\nonumber \\\nonumber
& - m_{\tilde{e}_R}^2 \tilde{e}_R^* \tilde{e}_R - (\epsilon_{ij} [y_e A_e H_d^i
\tilde{L}^j \tilde{e}_R^* + y_d
A_d H_d^i \tilde{Q}^j \tilde{d}_R^* - y_u A_u H_u^i \tilde{Q}^j
\tilde{u}_R^*] + h.c.) \\
& -\frac{1}{2}(M_1 \tilde{B}\tilde{B} + M_2
\tilde{W}_i\tilde{W}_i + M_3 \tilde{G}\tilde{G} + h.c.)\\ \nonumber
&- m_S^2 |S|^2 +
(\epsilon_{ij} \lambda 
A_\lambda S H_d^i H_u^j - \frac{1}{3} \kappa
A_\kappa S^3 + \hc) \;,
\label{eq:softmssm}
\end{align}
where exemplary for the first generation $\tilde{Q}=(\tilde{u}_L,\tilde{d}_L)^T$ and
$\tilde{L}=(\tilde{\nu}_L,\tilde{e}_L)^T$ denote the complex scalar
components of the corresponding quark and lepton superfields. Working in
the CP-violating NMSSM the soft SUSY breaking trilinear couplings $A_x$
($x=\lambda,\kappa,d,u,e$) and the gaugino mass parameters $M_k$
($k=1,2,3$) of the bino, wino and gluino fields
$\tilde{B},\tilde{W}_i$ ($i=1,2,3$) and $\tilde{G}$ 
are taken to be complex. By exploiting the $R$-symmetry either $M_1$
or $M_2$ can chosen to be real. The soft SUSY breaking mass parameters of the
scalar fields, $m_X^2$ ($X=S, H_d, H_u,
\tilde{Q},\tilde{u}_R,\tilde{d}_R,\tilde{L},\tilde{e}_R$) are real. A
sum over all three quark and lepton generations is implicit.  

\subsection{The Higgs Sector at Tree Level \label{sec:higgssector}}
From the superpotential, the soft SUSY breaking terms and the $D$-term
contributions the Higgs potential is obtained as,
\beq
V_{H}  &=& (|\lambda S|^2 + m_{H_d}^2)H_{d,i}^* H_{d,i}+ (|\lambda S|^2
+ m_{H_u}^2)H_{u,i}^* H_{u,i} +m_S^2 |S|^2 \nonumber \\
&& + \fr18 (g_2^2+g_1^{2})(H_{d,i}^* H_{d,i}-H_{u,i}^* H_{u,i} )^2
+\fr12g_2^2|H_{d,i}^* H_{u,i}|^2 \label{eq:higgspotential} \\ 
&&   + |-\epsilon^{ij} \lambda  H_{d,i}  H_{u,j} + \kappa S^2 |^2+
\big[-\epsilon^{ij}\lambda A_\lambda S   H_{d,i}  H_{u,j}  +\fr13 \kappa
A_{\kappa} S^3+\hc \big] \,,
\nonumber
\eeq
where $g_1$ and $g_2$ denote the $U(1)_Y$ and $SU(2)_L$ gauge
couplings, respectively. 
The expansion of the two Higgs doublets and the singlet field about their
vacuum expectation values, $v_d, v_u$ and $v_s$, introduces two
additional phases,  $\varphi_u$ and $\varphi_s$,
\beq
H_d =
 \bpmatrix \fr{1}{\sqrt 2}(v_d + h_d +i a_d)\\ h_d^- \epmatrix,\quad
H_u = e^{i\varphi_u}\bpmatrix
h_u^+ \\ \fr{1}{\sqrt 2}(v_u + h_u +i a_u)\epmatrix,\quad
S= \fr{e^{i\varphi_s}}{\sqrt 2}(v_s + h_s +ia_s).~
\label{eq:Higgs_decomposition} 
\eeq
The phase $\varphi_u$ enters the top quark mass. In order to keep
the top Yukawa coupling real, we absorb this phase into the
left-handed and right-handed top fields by replacing 
\be  
t_L \to e^{-i\varphi_u/2}\,t_L \quad \mbox{and} \quad t_R \to
e^{i\varphi_u/2}\,t_R\,. \label{eq:topfieldrephase}
\ee
This affects all couplings involving one top quark. Substituting
\eqref{eq:Higgs_decomposition} into 
\eqref{eq:higgspotential}, the Higgs potential can be cast into the form
\begin{align}
 V_H = & V_H^{\mbox{\scriptsize const}}  +  t_{h_d} h_d + t_{h_u} h_u +  t_{h_s} h_s  +  t_{a_{d}} a_d+  t_{a_{u}} a_u
+  t_{a_{s}} a_s  \\ \non
&+ 
 \fr 12 \bpmatrix h_d,h_u,h_s,a_d,a_u,a_s \epmatrix  {\mathcal{M}_{\phi\phi}}
\bpmatrix  h_d\\h_u\\h_s \\a_d\\a_u\\a_s \epmatrix+ \bpmatrix h_d^+, h_u^+\epmatrix {\mathcal{M}_{h^+h^-}} \bpmatrix h_d^-\\ h_u^-\epmatrix
+V_H^{\phi^3,\phi^4} \,,
\label{eq:vhpot}
\end{align} 
with the tadpole coefficients  $t_{\phi}$ ($\phi=h_d,h_u,h_s,a_d,a_u,a_s$),
the $6\times 6$ mass matrix ${\mathcal{M}}_{\phi\phi}$ for the neutral
Higgs bosons and the $2\times 2$ mass matrix ${\mathcal{M}}_{h^+h^-}$ for
the charged Higgs bosons. 
The constant terms are summarized in 
$V_H^{\mbox{\scriptsize const}}$ and the trilinear and quartic Higgs interactions in
$V_H^{\phi^3, \phi^4}$. The explicit expressions for the tadpoles and
mass matrices $\mathcal{M}_{\phi\phi}$ and $\mathcal{M}_{h^+h^-}$ are
given in \bib{Graf:2012hh}. As they are rather lengthy we do not
repeat them here, but summarize their main features:
\begin{itemize}
\item At tree level, the tadpole coefficients vanish due to the
requirement of the Higgs potential taking its minimum 
at the VEVs $v_d,v_u$ and $v_s$. However, only five of the six minimum
conditions are actually linearly independent.
\item  The three phase combinations that appear in the tadpoles and the mass
matrices at tree level are given by
\bea
\varphi_x &=& \varphi_{A_\lambda}+ \varphi_{\lambda}+\varphi_{s}+\varphi_{u} \,,\\
\varphi_y &=& \varphi_{\kappa} -\varphi_{\lambda} + 2\varphi_{s}-\varphi_{u}\,,\\
\varphi_z &=& \varphi_{A_\kappa} + \varphi_{\kappa}+ 3\varphi_{s} \,.
\eea
At lowest order, two of them can be eliminated by exploiting the
minimization conditions $t_{a_d}=0$ and $t_{a_s}=0$. We choose 
$\varphi_x$ and $\varphi_z$ to be expressed in terms of $\varphi_y$,
so that all mass matrix elements mixing the CP-even and CP-odd
interaction states, ${\mathcal{M}}_{h_i a_j}$, are proportional to
$\sin \varphi_y$. This is the only CP-violating phase that occurs at
tree level in the Higgs sector.  
\item The transformation from the interaction states to the mass
  eigenstates is performed in two steps. First the would-be Goldstone
  boson field is separated via rotation by the matrix $\mathcal{R}^G$, 
  then the matrix $\mathcal{R}$ is used to rotate to the mass eigenstates,
\beq
(h_d,h_u,h_s,a,a_s, G)^T &=&  \mathcal{R}^G~(
h_d,h_u,h_s,a_d,a_u,a_s)^T\,, \non \\ 
(h_1,h_2,h_3,h_4,h_5, G)^T &=& \mathcal{R} ~(h_d,h_u,h_s,a,a_s,
G)^T\,, \label{eq:rotationtreelevel}
\eeq
with the diagonal mass matrix
\beq
\diag(m_{h_1}^2,m_{h_2}^2,m_{h_3}^2,m_{h_4}^2,m_{h_5}^2,0)&=&
\mathcal{R} \mathcal{M}_{hh} \mathcal{R}^T\,, \quad \mathcal{M}_{hh}=
\mathcal{R}^G\mathcal{M}_{\phi\phi}(\mathcal{R}^G)^T. 
\label{eq:massmatrix}
\eeq
The mass eigenstates $h_i$ ($i=1,...,5$) are ordered by ascending
mass, with the lightest mass given by $m_{h_1}$.
\item The tree-level mass of the charged Higgs boson reads
\be 
M_{H^\pm}^2 = M_W^2 +\fr{|\lambda|v_s}{s_{2\beta}}\bpmatrix
\sqrt{2}|A_{\lambda}|c_{\varphi_x}+ |\kappa|v_s c_{\varphi_y}
\epmatrix - \fr{|\lambda|^2v^2}{2}\,,
\ee
where here and in the following we use the short hand notations
$c_x=\cos x, s_x=\sin x$ and $t_x=\tan x$. The vacuum expectation
value $v\approx 246$~GeV is related to $v_u$ and $v_d$ through
$v^2=v_d^2+v_u^2$.
\item The MSSM limit is obtained by $\lambda, \kappa \to 0$ and
  keeping the parameter $|\mu_{\text{eff}}|=|\lambda| v_s /\sqrt{2}$
  as well as $A_\lambda$ and $A_\kappa$ fixed. In this limit the
  mixing between the singlet and the doublet fields goes to zero. 
\end{itemize}
The set of independent parameters entering the Higgs potential at
tree level is chosen to be 
\be 
t_{h_d},t_{h_u},t_{h_s},t_{a_d},t_{a_s},M_{H^\pm}^2,v,s_{\theta_W},
e,\tan\beta,|\lambda|,v_s,|\kappa|,\text{Re}A_{\kappa},s_{\varphi_y}\,.
\ee
There are several changes
with respect to the parameter set chosen in \bib{Graf:2012hh}. Here we 
use $v$ and $s_{\theta_W}$ instead of $M_W$ and $M_Z$, since this is
more convenient for the computation of the order ${\cal O}(\al_t\al_s)$
corrections to the Higgs boson masses, in which we work in the gaugeless
limit, \ie $e=0$ and $M_W=M_Z=0$ but $v\ne 0$ and $s_{\theta_W}\ne
0$. Furthermore the real part of $A_\kappa$ is considered rather 
than the absolute value. In accordance with the SUSY Les
Houches Accord (SLHA) \cite{Skands:2003cj,Allanach:2008qq} conventions
we regard the real part as an input parameter and use the tadpole
conditions to eliminate the imaginary part of $A_\kappa$.
For $\lambda$ and $\kappa$ this distinction is not necessary, since
both the real and imaginary parts are given in the SLHA convention 
and can be related to the respective absolute values and phases.

\subsection{The Quark and Squark  Sector}
\label{sec:colorsector}
The two-loop diagrams of the order ${\cal O}(\al_t\al_s)$ 
contain coloured particles like top quark, stop, gluon and gluino in the 
self-energies of the neutral Higgs bosons and additionally bottom
quark and sbottom in the charged Higgs self-energy.  The stop sector
of the complex NMSSM differs from the one of the MSSM due to the
appearance of the new complex phase $\varphi_u$.  \s

In the gaugeless approximation $e\to 0$, the stop mass matrix reads 
\begin{equation}
\mathcal{M}_{\tilde t}=
\bpmatrix
 m_{\tilde{Q}_3}^2+m_t^2 & m_t \left(A_t^* e^{-i\varphi_u}-\fr{\mueff}{\tan\beta}\right) \\[2mm]
m_t \left(A_t e^{i\varphi_u}- \fr{\mueff^*}{\tan\beta}\right) & m_{\tilde{t}_R}^2+m_t^2
\epmatrix\,,
\end{equation}
where the effective higgsino mixing parameter  
\beq
\mueff= \fr{\lambda v_s e^{i\varphi_s}}{\sqrt{2}} \label{eq:mueff}
\eeq
has been introduced. The matrix is diagonalized by a unitary matrix
$\mathcal{U}_{\tilde t}$, rotating the interaction states $\tilde{t}_L$
and $\tilde{t}_R$ to the mass eigenstates $\tilde{t}_1$ and $\tilde{t}_2$,
\bea
  (\tilde t_1,\tilde t_2)^T &=& \mathcal{U}_{\tilde t}~(\tilde
  t_L,\tilde t_R)^T ,\\
 \text{diag}(m_{\tilde t_1}^2,m_{\tilde t_2}^2)&=&\mathcal{U}_{\tilde t}~\mathcal{M}_{\tilde t}~\mathcal{U}_{\tilde t}^\dagger\,.
\eea
In the two-loop diagrams of the charged Higgs self-energy we treat the
bottom quark as massless, {\it i.e.}~$m_b=0$. Consequently the left- and
right-handed sbottom states do not mix and only the left-handed
sbottom with a mass of $m_{\tilde{Q}_3}$ contributes. Summarizing, the
set of independent parameters entering the top/stop and bottom/sbottom
sector is chosen to be   
\be 
m_t, \; m_{\tilde{Q}_3}, \; m_{\tilde t_R} \quad \mbox{and} \quad A_t \;.
\label{eq:stopparset}
\ee 
With this parameter choice for the mass matrix in the interaction
basis the rotation matrix $\mathcal{U}_{\tilde t}$ does not need to be
renormalized. This is the same approach as used in the Higgs sector,
where we do not renormalize the rotation matrices either. \s

The parameters in Eq.~(\ref{eq:stopparset}) are
renormalized at ${\cal O}(\al_s)$. 
The renormalization can be performed in the on-shell (OS)
\cite{Heinemeyer:2007aq,Heinemeyer:2010mm} or $\DRb$
scheme. For the values of the input parameters we follow the SLHA in
which the top quark mass is taken to be the pole mass whereas the soft
SUSY breaking masses and trilinear couplings are 
understood as $\overline{\mbox{DR}}$ parameters evaluated at the
renormalization scale $\mu_R = M_{\text{SUSY}}$. The latter will be
specified in the numerical analysis in Section~\ref{sec:numerical}. 
For the numerical evaluation of the two-loop corrected
Higgs boson masses in {\tt NMSSMCALC} 
both renormalization schemes have been implemented,
and the user has the choice to switch from the default $\DRb$ scheme
to the OS scheme by setting the corresponding flag in the input file. 
The translation between the two schemes is performed consistently both in the
counterterm part and at the level of the input parameters. 
The OS and $\overline{\mbox{DR}}$ counterterms for any of the
parameters $X \equiv m_t, m_{\tilde{Q}_3}, m_{\tilde t_R}$ and $A_t$ can be
expanded in terms of the  dimensional regularization parameter $D= 4 -
2\epsilon$ as 
\bea 
\delta X^{\OS} &=& \fr{1}{\epsilon} \delta X_{\text{pole}} +  \delta
X_{\text{fin}} \, , \label{eq:OScounterterm}\\
 \delta X^{\DRb} &=& \fr{1}{\epsilon} \delta X_{\text{pole}}\,.  
\label{eq:DRbarcounterterm}
\eea
This fixes the relation between the counterterms in the two
schemes. Our definition of the parameters in the OS scheme
deliberately does not take into account any terms that are
proportional to $\epsilon$, {\it i.e.}~$\epsilon \delta
X_\epsilon$. Of course one could also choose to include such terms,
that would then manifest themselves as additional finite contributions, due to
the counterterm inserted diagrams multiplying $1/\epsilon$ terms
from the one-loop functions with the $\epsilon$ parts of the
counterterms. We apply our thus defined OS scheme 
consistently throughout the whole calculation. 
Choosing the input according to the SLHA and the $\overline{\mbox{DR}}$
scheme as default renormalization scheme, first the $m_t^{\DRb}$ has to be
computed from the corresponding OS parameter $m_t^{\OS}$ as described
in Appendix~\ref{app:mtoprun}. When switching to the OS scheme the
translation of the parameters $m_{\tilde{Q}_3}^2, m_{\tilde t_R}^2$ and
$A_t$ from the $\overline{\mbox{DR}}$ scheme to the OS scheme is
performed by applying
\beq
A_t^{(\OS)} = A_t^{(\DRb)} - \delta A_t^{\text{fin}} \;, \label{eq:atfin}\\
(m_{\tilde Q_L}^2)^{(\OS)}  = (m_{\tilde Q_L}^2)^{(\DRb)} - \delta 
(m_{\tilde Q_L}^2)^{\text{fin}} \;, \label{eq:qlfin}\\
(m_{\tilde t_R}^2)^{(\OS)}  = (m_{\tilde t_R}^2)^{(\DRb)} - \delta
(m_{\tilde t_R}^2)^{\text{fin}} \label{eq:trfin} \;.
\eeq
Note that we computed the finite counterterm parts in
Eqs.~(\ref{eq:atfin})-(\ref{eq:trfin}) with 
OS input parameters. Hence an iterative procedure is
required to obtain the OS parameters. The OS conditions for the
complex MSSM (s)quark sector, which 
is the same in the NMSSM, have been presented in
Refs.~\cite{Heinemeyer:2007aq} and \cite{Heinemeyer:2010mm}. For
completeness, we list here the expressions for the counterterms.
\begin{itemize}
\item The top mass counterterm reads
\begin{equation}
 \delta m_t=\frac{1}{2}\widetilde{\Re} \;\Big(m_t \Sigma_t^{VL}(m_t^2)+ m_t \Sigma_t^{VR}(m_t^2)+\Sigma_t^{SL}(m_t^2)+\Sigma_t^{SR}(m_t^2)\Big)\,,
\end{equation}
where $\widetilde{\Re}$ means that the real part is taken only for the
one-loop integral function but not for the parameters. The
unrenormalized top self-energy $\Sigma_t$ is decomposed as
\begin{equation}
 \Sigma_t(p^2)=\slashed p P_L \Sigma^{VL}_t(p^2)+\slashed p P_R \Sigma^{VR}_t(p^2)+P_L \Sigma^{SL}_t(p^2)+P_R \Sigma^{SR}_t(p^2)\,, 
\end{equation}
with the left- and right-handed projectors $P_{L/R}=(1\mp\gamma_5)/2$.
\item The counterterm of the trilinear top coupling is given by
\beq
  \delta A_t=\frac{e^{-i\varphi_u}}{m_t}&& \hspace*{-0.5cm}
\left[\mathcal{U}_{\tilde 
t_{11}}\mathcal{U}_{\tilde t_{12}}^*(\delta m_{\tilde t_1}^2-\delta 
m_{\tilde t_2}^2)+
\mathcal{U}_{\tilde t_{11}}\mathcal{U}_{\tilde t_{22}}^* (\delta Y)^*+
\mathcal{U}_{\tilde t_{21}}\mathcal{U}_{\tilde t_{12}}^* \delta Y
\right. \nonumber \\ && \hspace*{-0.5cm}
\left. 
-\left(A_t e^{i \varphi_u}-\fr{\mueff^*}{\tan\beta}\right) \delta m_t
\right]\,,
\eeq
where
\bea
\delta m_{\tilde t_1}^2 &=&\Sigma_{\tilde t_1 \tilde t_1}(m_{\tilde t_1}^2)\,,\\
\delta m_{\tilde t_2}^2 &=&\Sigma_{\tilde t_2 \tilde t_2}(m_{\tilde t_2}^2)\,,\\
 \delta Y &\equiv& \big[\mathcal{U}_{\tilde t}\delta\mathcal{M}_{\tilde t}\mathcal{U}_{\tilde t}^\dagger\big]_{12}=\big[\mathcal{U}_{\tilde t}\delta\mathcal{M}_{\tilde t}\mathcal{U}_{\tilde t}^\dagger\big]^*_{21}=\frac{1}{2}\widetilde\Re\;\Big(\Sigma_{\tilde t_1^* \tilde t_2^*}(m_{\tilde t_1}^2)+\Sigma_{\tilde t_1^* \tilde t_2^*}(m_{\tilde t_2}^2)\Big)\,.
\eea
We denote by $\Sigma_{\tilde t_i \tilde t_j}$ the unrenormalized 
self-energy for the $\tilde t_i \rightarrow \tilde t_j$ transition. 
\item The counterterm for the soft SUSY breaking left-handed squark
  mass parameter reads
\bea
 \delta m_{\tilde Q_L}^2=|\mathcal{U}_{\tilde t_{11}}|^2\delta m_{\tilde t_1}^2+|\mathcal{U}_{\tilde t_{12}}|^2\delta m_{\tilde t_2}^2-\mathcal{U}_{\tilde t_{22}}\mathcal{U}_{\tilde t_{12}}^*\delta Y-\mathcal{U}_{\tilde t_{12}}\mathcal{U}_{\tilde t_{22}}^* (\delta Y)^*
-2m_t\delta m_t\,.
\eea
\item Finally, the counterterm for the soft SUSY breaking right-handed
  stop mass parameter is given by
\beq
 \delta m_{\tilde t_R}^2=|\mathcal{U}_{\tilde t_{21}}|^2\delta m_{\tilde t_1}^2+|\mathcal{U}_{\tilde t_{22}}|^2\delta m_{\tilde t_2}^2-\mathcal{U}_{\tilde t_{11}}\mathcal{U}_{\tilde t_{21}}^* (\delta Y)^* -\mathcal{U}_{\tilde t_{21}}\mathcal{U}_{\tilde t_{11}}^* \delta Y-2m_t\delta m_t\,.
\eea
\end{itemize}

To complete this subsection, we present the Lagrangians containing the
charged Higgs coupling to a top-bottom pair and the top-stop-gluino
coupling as well as the charged $W$ boson coupling to top and
bottom quark. These are affected by the absorption of the phase related to 
$v_u$ in the top quark field, {\it cf.}~\eqref{eq:topfieldrephase},
\bea
{\cal L}_{t\bar{b}H^-}&=& y_t \cos\beta e^{-i\fr{\varphi_u}{2}}
\bar{b} P_R t H^- +\hc\;, \\
{\cal L}_{t\tilde{t}\tilde{g}}&=& \sqrt{2} g_s\bar{\tilde g}^a\left( -T^a_{jk} \mathcal{U}_{\tilde t_{i1}} e^{-i \fr{\varphi_{M_3}+\varphi_{u}}{2}}P_L
+ T^a_{jk} \mathcal{U}_{\tilde t_{i2}} e^{i
  \fr{\varphi_{M_3}+\varphi_{u}}{2}}P_R \right) t^k
\tilde{t}^{j*}_i+\hc\;, \\
{\cal L}_{t\bar{b}W^-}&=&-\frac{g_2}{\sqrt{2}}e^{-i\frac{\varphi_u}{2}} 
\bar{b}\gamma^{\mu}P_L t W^-_{\mu} +\hc \;, 
\eea 
 where $y_t = \sqrt{2} m_t/(v\sin\beta)$, $i=1,2$ denotes the sfermion
 mass eigenstate, $j,k=1,2,3$ the $SU(3)$ color indices, $g_s$ the
 strong coupling and $T^a$ ($a=1,...,8$) the generators of $SU(3)$. 

\section{The NMSSM Higgs Boson Masses at Order ${\cal
    O}(\al_t\al_s)$ \label{sec:calculation}}
In the Feynman diagrammatic approach, the two-loop Higgs masses are
obtained by determining the poles of the propagators, which is
equivalent to the calculation of the zeros of the determinant of the
two-point function $\hat\Gamma(p^2)$,
\be
\text{Det}\left(\hat\Gamma(p^2)\right)=0\;,\quad \text{with} \quad
\left(\hat\Gamma(p^2)\right)_{ij}=i\delta_{ij}\left(p^2-m_{h_i}^2\right)
+i\hat\Sigma_{ij}(p^2) 
\; , \qquad i,j=1...5 \;,
\ee
where $m_{h_i}$ are the tree-level masses and $\hat\Sigma_{ij}(p^2)$
is the renormalized self-energy of the $h_i\rightarrow h_j$ transition
at $p^2$. Note that $h_{i/j}$ denote the tree-level mass eigenstates. 
We have neglected the higher order corrections due to the mixing of 
the Goldstone boson with the remaining neutral Higgs bosons. This mixing has
been  verified numerically to be negligible. For the evaluation of the
loop-corrected Higgs masses and Higgs mixing matrix, we follow the
numerical procedure given in
Refs.\cite{Ender:2011qh,Graf:2012hh}. \s

The renormalized self-energies of the Higgs bosons, $\hat\Sigma_{ij}$,  
contain one-loop and two-loop contributions, which are labeled with
the superscript (1) and (2), respectively,
\be
\label{eq:selfsum}
\hat\Sigma_{ij}(p^2)=\hat\Sigma_{ij}^{(1)}(p^2)+\hat\Sigma_{ij}^{(2)}(0)\;.
\ee 
The one-loop renormalized self-energies have been discussed in detail
in Refs.~\cite{Ender:2011qh} and \cite{Graf:2012hh}. Here we concentrate
only on the two-loop parts, $\hat\Sigma_{ij}^{(2)}(0)$. They are
evaluated at vanishing external momentum $p^2 =0$ and can be decomposed as 
\begin{equation}
 \hat\Sigma^{(2)}_{h_ih_j}(0)=\Sigma^{(2)}_{h_ih_j}(0)-\frac{1}{2}\Big[\mathcal{R}
\Big(( \wavetwo)^\dagger\mathcal{M}_{hh}+ \mathcal{M}_{hh}
\wavetwo\Big)
\mathcal{R}^T\Big]_{ij}-\Big(\mathcal{R}\deltatwo
\mathcal{M}_{hh}\mathcal{R}^T\Big)_{ij}\,, 
\label{Eq:Eren2loop}
\end{equation}
where the first term, $\Sigma^{(2)}_{h_ih_j}$, denotes the
unrenormalized self-energy,  the second term contains the wave
function renormalization constants with 
\be 
\wavetwo=\diag (\deltatwo Z_{H_d},\deltatwo Z_{H_u},\deltatwo
Z_{S},s_\beta^2\deltatwo Z_{H_d}+c_\beta^2\deltatwo Z_{H_u},\deltatwo
Z_{S})\,, 
\ee  
and the third term includes the two-loop counterterm mass matrix
$\deltatwo \mathcal{M}_{hh}$. The neutral Higgs mass matrix
$\mathcal{M}_{hh}$ and the rotation matrix $\mathcal{R}$ correspond to
the ones defined in Eqs.~(\ref{eq:rotationtreelevel}) and
(\ref{eq:massmatrix}) after dropping the Goldstone component. The
counterterm constants appearing in \eqref{Eq:Eren2loop} will be discussed
in detail in \ssect{sec:fixcount}. 

\subsection{The Unrenormalized Self-Energies of the Neutral Higgs
  Bosons} 
\begin{figure}[tb]
\begin{center}
 \includegraphics[width=14cm]{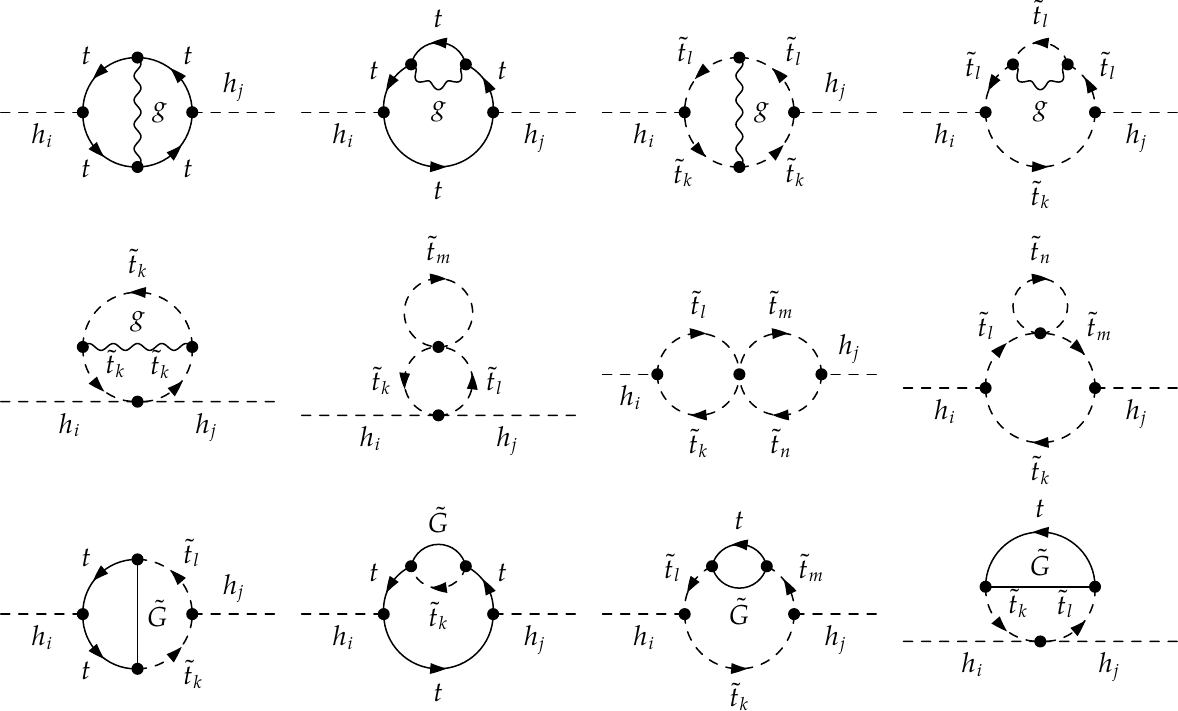}
\end{center}
\caption{Sample diagrams of genuine two-loop corrections contributing
  to the neutral Higgs boson self-energies at $\order$, with tops
  ($t$), stops ($\tilde t_{1},\tilde t_2$), gluons ($g$) and gluinos ($\tilde G$) in the
  loops and $k,l,m,n=1,2$, $i,j=1,...,5$.}
\label{Fig_neutral}
\end{figure}
\begin{figure}[tb]
\begin{center}
  \includegraphics[width=15cm]{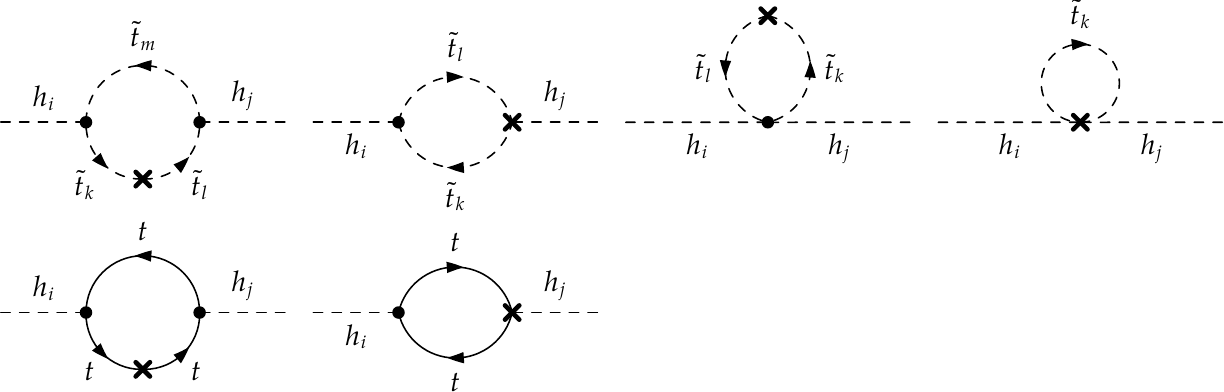}
\end{center}
\caption{Examples of counterterm inserted diagrams contributing to the
  neutral Higgs boson self-energies.} 
\label{Fig_neutralb}
\end{figure}

The  unrenormalized self-energies of the transitions $h_i\rightarrow
h_j$ consist of the contributions from genuine two-loop diagrams  and
from counterterm inserted one-loop diagrams. The genuine two-loop
diagrams must contain either a gluon or gluino or four-stop
couplings. Some example diagrams are presented in \figref{Fig_neutral}.
After performing the tensor reduction of these diagrams at zero
external momentum, an expression in terms of either one two-loop vacuum
integral or of products of two one-loop vacuum integrals is
obtained. The counterterm inserted one-loop diagrams, which are shown
in \figref{Fig_neutralb}, contain either coupling-type counterterms or
propagator-type  counterterms of top quarks and stops. The  set of
counterterms involved in these diagrams has been discussed in 
\ssect{sec:colorsector}. For the evaluation of these counterterms also
the one-loop two-point functions with full momentum dependence are
needed. The one-loop and the two-loop master integrals have to be expanded
in terms of the dimensional regularization parameter $D= 4 -
2\epsilon$. The one-loop one-point and two-point functions have
been defined in \cite{'tHooft:1978xw,Nierste:1992wg}. For the two-loop
vacuum functions we use the existing results in
\cite{Davydychev:1992mt,Ford:1992pn,Scharf:1993ds,Weiglein:1993hd,Berends:1994ed,Martin:2001vx,Martin:2005qm}. Inserting
these expansions into the two-loop expressions, we can easily
extract the coefficients of the double pole, single pole and finite
parts. After gaining such expressions also for the counterterms of the
relevant parameters we can explicitly check the cancellation of the UV
divergences. 

\subsection{The Counterterms}\label{sec:fixcount}

When calculating the $\order$ corrections, we employ the gaugeless limit \ie $e\to 0$. This leads to  the independence of the Higgs potential on $s_{\theta_W}$. Therefore  
we restrict ourselves to a new set of independent parameters entering
the Higgs potential at order ${\cal O}(\al_t\al_s)$, 
\be 
t_{h_d},t_{h_u},t_{h_s},t_{a_d},t_{a_s},M_{H^\pm}^2,v,\tan\beta,|\lambda|,v_s,|\kappa|,\text{Re}A_{\kappa},s_{\varphi_y}\,. 
\ee
In order to obtain a UV-finite result, these parameters need to be
renormalized. The parameters are replaced by the renormalized ones and
the corresponding counterterms according to 
\begin{align} 
t_{\phi} &\rightarrow t_{\phi} +\deltaone t_{\phi}  +\deltatwo 
t_{\phi} \qquad\qquad\quad \text{with}~~\phi=h_d,h_u,h_s,a_d,a_s
\;,\label{eq:tphi}\\
M_{H^\pm}^2 &\rightarrow M_{H^\pm}^2+  \deltaone M_{H^\pm}^2+  \deltatwo M_{H^\pm}^2\;,\\
v &\rightarrow v + \deltaone v +  \deltatwo v\;,\\
\tan\beta &\rightarrow \tan\beta  +  \deltaone \tan\beta+  \deltatwo \tan\beta\;,\\
v_s &\rightarrow v_s+  \deltaone v_s +  \deltatwo v_s\;,\\
|\lambda| &\rightarrow |\lambda|+  \deltaone |\lambda| +  \deltatwo |\lambda|\;,\\
|\kappa| &\rightarrow |\kappa|+  \deltaone |\kappa|+  \deltatwo |\kappa|\;,\\
\text{Re}A_{\kappa} &\rightarrow \text{Re}A_{\kappa}+  \deltaone \text{Re}A_{\kappa} +  \deltatwo \text{Re}A_{\kappa}\;,\\
s_{\varphi_y} &\rightarrow s_{\varphi_y}+ \deltaone s_{\varphi_y} +
\deltatwo s_{\varphi_y}\;,
\label{eq:sphiy}
\end{align}
where the superscript $(n)$ denotes the $n$-loop level. The one-loop
counterterms are of course not of order ${\cal O}(\alpha_t\alpha_s)$ and
have been defined explicitly in \bib{Graf:2012hh}. Therefore we 
restrict ourselves here to the discussion of the two-loop
counterterms only. To ensure consistency of our one- and two-loop
corrections we apply the same mixed $\DRb$-on-shell renormalization
scheme as in \bib{Graf:2012hh}, in particular,\footnote{In a slight
  abuse of the language we use the expression OS, although we put the
  external momenta to zero.}  
\be
\underbrace{ t_{h_d},t_{h_u},t_{h_s},t_{a_d},t_{a_s},M_{H^\pm}^2,v}_{\mbox{on-shell
 scheme}}, 
\underbrace{\tan\beta,|\lambda|,v_s,|\kappa|,\text{Re}A_{\kappa},s_{\varphi_y}}_{\overline{\mbox{DR}} \mbox{ scheme}}\,. 
\label{eq:mixedcond}
\ee
Inserting the replacements given in Eqs.~(\ref{eq:tphi})-(\ref{eq:sphiy}) in the mass matrix, the counterterm  matrix for the
neutral Higgs mass matrix $\mathcal{M}_{hh}$ is obtained,
\be 
\mathcal{M}_{hh} \rightarrow  \mathcal{M}_{hh}
+\deltaone\mathcal{M}_{hh}   +\deltatwo\mathcal{M}_{hh} \;.
\ee
In \appen{sec:dMH2loop}, we give the explicit expressions of
$\deltatwo\mathcal{M}_{hh}$ in terms of  all OS parameter
counterterms.  \s

In addition to the set of independent parameters of
Eq.~(\ref{eq:mixedcond}), the Higgs field 
wave functions need to be renormalized. 
The renormalization constants for the doublet and singlet fields are
introduced before rotating to the mass eigenstates as 
\bea 
H_d&\rightarrow& (1+ \fr{1}2 \deltaone Z_{H_d}+ \fr{1}2 \deltatwo
Z_{H_d})H_d \;,\\
H_u&\rightarrow& (1+ \fr{1}2 \deltaone Z_{H_u}+ \fr{1}2 \deltatwo
Z_{H_u})H_u \;,\\
S&\rightarrow& (1+ \fr{1}2 \deltaone Z_{S}+ \fr{1}2 \deltatwo
Z_{S})S \;. 
\eea 
The counterterms for the renormalized parameters and the field
renormalization constants are fixed via the renormalization
conditions, listed in the following:
\begin{itemize}
\item Analogously to the one-loop calculation the field
  renormalization constants are given via $\DRb$ conditions defined as
\bea
\deltatwo Z_{H_d}= -\left.\fr{\partial \Sigma^{(2)}_{h_dh_d}}{\partial p^2}\right|_{\text{div}}(p^2\to 0)\,,\\
\deltatwo Z_{H_u}= -\left.\fr{\partial \Sigma^{(2)}_{h_uh_u}}{\partial p^2}\right|_{\text{div}}(p^2\to 0)\,,\\
\deltatwo Z_{S}= -\left.\fr{\partial \Sigma^{(2)}_{h_sh_s}}{\partial p^2}\right|_{\text{div}}(p^2\to 0)\,,
\eea 
where the subscript '$\text{div}$' denotes the divergent part. It
turns out that at order ${\cal O}(\al_t\al_s)$ $\deltatwo Z_{H_d}$ and
$\deltatwo Z_{S}$ are zero.

Using $\DRb$ renormalization in the top/stop sector the non-vanishing field renormalization
constant $\deltatwo Z_{H_u}$ is given by\footnote{Please, note that
  the superscript $\DRb$ on $\deltatwo Z_{H_u}$ is supposed to
  indicate that the $\DRb$ top mass is used in the calculation as
  opposed to the pole mass, in which case we will write $\deltatwo
  Z_{H_u}^{\OS}$. This should not be confused with the use of an
  on-shell condition for the field renormalization constant itself,
  \ie the inclusion of a finite part.} 
\be
\deltatwo Z_{H_u}^{\DRb} =\fr{\al_s(m_t^2)^{\DRb}}{8 \pi^2
  v^2\sin^2\beta }\left( \fr{1}{\epsilon^2} -\fr{1}{\epsilon}\right),
\ee
which is in agreement with the result as obtained in the MSSM
\cite{Degrassi:2014pfa}. If instead one uses on-shell renormalization
for the top mass, the counterterm inserted one-loop diagrams will lead
to an additional contribution to the field renormalization constant, which is then
\be 
\deltatwo Z_{H_u}^\OS=\fr{\al_s(m_t^2)^\OS}{8 \pi^2 v^2\sin^2\beta }\left( \fr{1}{\epsilon^2}
  -\fr{1}{\epsilon}\right)-\frac{3}{4\pi^2}\frac{m_t^\OS (\delta
  m_t)_{\text{fin}}}{v^2\sin^2\beta}\frac{1}{\epsilon} \,,
\ee
where $(\delta m_t)_{\text{fin}}$ is the finite part of the top mass
counterterm as defined in \eqref{eq:OScounterterm} and which is given by
\begin{align}
  (\delta m_t)_{\text{fin}}=
& \fr{\alpha_sm_t}{3\pi}\Big[3\log\Big(\frac{m_t^2}{\mu_R^2}\Big)-5\Big] 
+ dm_t \;.
  \end{align}
Here $dm_t$ is the SUSY-QCD correction given in
\eqref{eq:susyqcd_dmt}. However, we would like to point out that the 
complete wave function renormalization constant is the same up to
higher orders and independent of the renormalization scheme used for the
top mass. This can easily be seen, when looking at the sum of the one-
and two-loop counterterm $\delta Z_{H_u}$. At one-loop level, if one
takes only the top/stop contribution then $\deltaone Z_{H_d}$ and
$\deltaone Z_{S}$ are also zero and 
\be \deltaone Z_{H_u} = -\frac{3 m_t^2}{8 \pi^2v^2 \sin^2\beta}\frac
1\epsilon \;.
\ee
Hence, the sum of the one- and two-loop contribution is given by
\begin{align}
\delta Z_{H_u}^{\DRb}&=\underbrace{-\frac{3(m_t^2)^{\DRb}}{8 \pi^2v^2 \sin^2\beta}\frac 1\epsilon}_{\text{one-loop}}+\underbrace{\fr{\al_s(m_t^2)^{\DRb}}{8 \pi^2 v^2\sin^2\beta }\left( \fr{1}{\epsilon^2}
  -\fr{1}{\epsilon}\right)}_{\text{two-loop}},\\
\delta Z_{H_u}^{\OS}&=\underbrace{-\frac{3(m_t^2)^{\OS}}{8 \pi^2v^2 \sin^2\beta}\frac 1\epsilon}_{\text{one-loop}}+\underbrace{\fr{\al_s(m_t^2)^{\OS}}{8 \pi^2 v^2\sin^2\beta }\left( \fr{1}{\epsilon^2}
  -\fr{1}{\epsilon}\right)-\frac{3}{4\pi^2}\frac{m_t^\OS (\delta
  m_t)_{\text{fin}}}{v^2\sin^2\beta}\frac{1}{\epsilon}}_{\text{two-loop}}.
\end{align}
Taking into account the relation $m_t^{\DRb}=m_t^\OS+(\delta
m_t)_{\text{fin}}$, it is evident that $\delta Z_{H_u}^{\DRb}$ and
$\delta Z_{H_u}^{\OS}$ agree up to higher orders\footnote{The
  inclusion of the terms proportional to $\epsilon$ in the OS
  counterterm of $\delta m_t$ destroys this equality and would entail
  the conversion of further input parameters to match the two schemes.}.
\item The renormalization conditions for the tadpoles are chosen such
  that the minimum of the Higgs potential does not change when it receives
  two-loop corrections, leading to
\be 
\deltatwo t_{\phi}= t_{\phi}^{(2)}\;, \quad
\phi=(h_d,h_u,h_s,a_d,a_s) \;. 
\ee 
Sample two-loop tadpole and counterterm inserted tadpole diagrams are shown in
Fig.~\ref{fig:tad}.
\begin{figure}[t]
\begin{center}
 \includegraphics[width=15cm]{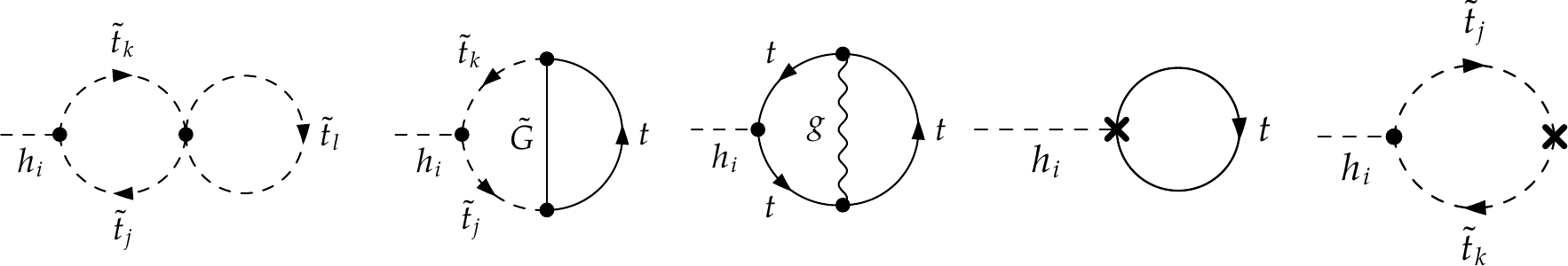}
\end{center}
\caption{Sample two-loop tadpole and counterterm inserted tadpole
  diagrams at $\order$ for the neutral Higgs bosons.} 
\label{fig:tad}
\end{figure}

\item The charged Higgs boson mass is defined as an OS parameter. Hence, 
\begin{align}
 \deltatwo M_{H^\pm}^2&=\Sigma_{H^\pm}^{(2)}(0)-M_{H^\pm}^2(\cos^2\!\beta \deltatwo Z_{H_u}+\sin^2\!\beta \deltatwo Z_{H_d})\,.\label{eq:dMHpm2}
\end{align}
The two-loop corrections to the mass of the charged Higgs boson are
also calculated at vanishing external momentum. Therefore the counterterm
for the mass of the charged Higgs boson is not solely fixed by the 
unrenormalized self-energy, but is also related to the field renormalization
constants.
Note, however, that this of course does not affect the finite
part. Some examples of two-loop diagrams contributing to the
unrenormalized self-energy of 
the charged Higgs boson are shown in Fig.~\ref{fig:chaHig} (upper row). In addition
to top quarks and squarks, gluons and gluinos also bottom quarks and
squarks appear in the loops. We perform our calculation in the limit
of vanishing bottom mass and neglect the $D$-term in the sbottom mass matrix. 
Therefore the left- and right-handed sbottoms do not mix, and only the
left-handed sbottom contributes. The counterterm of the
soft SUSY breaking left-handed squark mass parameter needed in the
calculation of the one-loop inserted counterterm diagrams has been
given in \ssect{sec:colorsector}. Some example diagrams are given in
Fig.~\ref{fig:chaHig} (lower row). 
\begin{figure}[t]
\begin{center}
 \includegraphics[width=15cm]{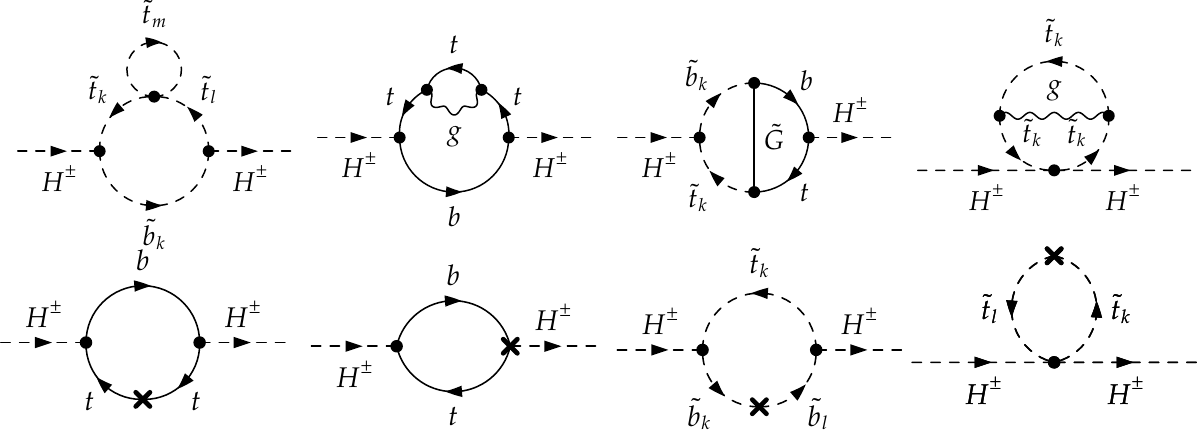}
\end{center}
\caption{Sample two-loop and counterterm inserted diagrams at
  $\order$ contributing to the charged Higgs boson self-energy.} 
\label{fig:chaHig}
\end{figure}

\begin{figure}[b]
\begin{center}
 \includegraphics[width=11cm]{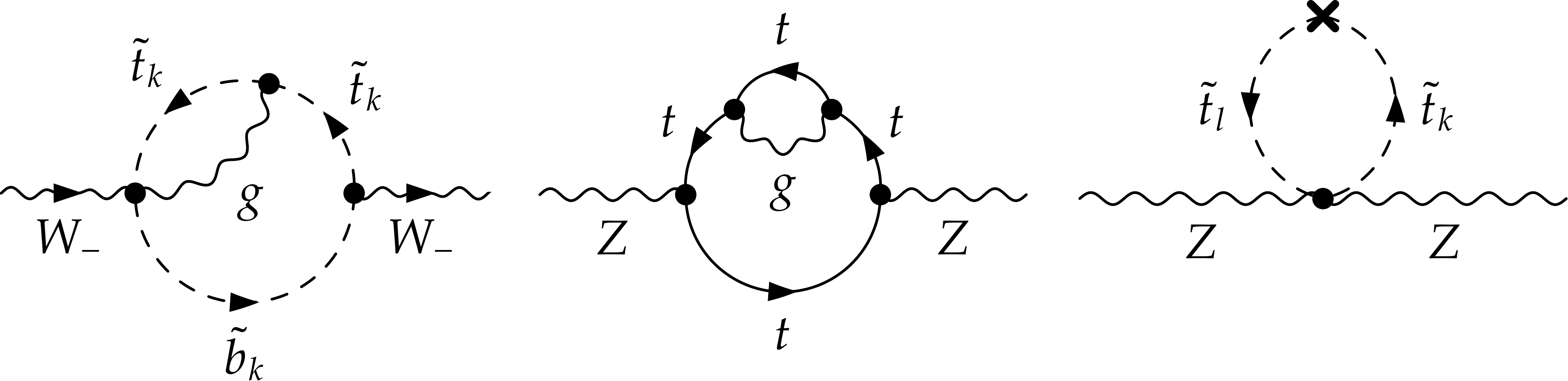}
\end{center}
\caption{Sample two-loop and counterterm inserted diagrams at
  $\order$ contributing to the $W$, respectively $Z$ boson self-energies.} 
\label{fig:vectorbosonself}
\end{figure}
\item The counterterm $\deltatwo v/v$ is taken to be an OS parameter
  and hence
\begin{equation}
\fr{\deltatwo v}{v}=\fr{c_{\theta_W}^2}{2 s_{\theta_W}^2}\left(
  \fr{\deltatwo M_Z^2}{M_Z^2} -\fr{\deltatwo M_W^2}{M_W^2} \right) +
\fr{\deltatwo M_W^2}{2M_W^2} \label{eq:dvexpression} \;,
\end{equation}
where
\begin{equation}
\fr{\deltatwo M_W^2}{M_W^2}=\fr{\Sigma_{W}^{T,(2)}(0)}{M_W^2} \qquad \text{and } \qquad \fr{\deltatwo M_Z^2}{M_Z^2}=\fr{\Sigma_{Z} ^{T,(2)}(0)}{M_Z^2}\,.
\end{equation}
Here $\Sigma^T_{V}(0)$ ($V=W,Z$) is the transverse part of the
unrenormalized vector boson self-energy. In the zero momentum
approximation, the transverse part relates to the one-particle irreducible two-point function as (we follow the convention in \texttt{FeynArts} \cite{Hahn:2000kx}),
\be 
\Gamma^{\mu\nu,(2)}_V(0)= -g^{\mu\nu} \Sigma^{T,(2)}_V(0)  \,.  
\ee
One should keep in mind that $\deltatwo M_V^2$ and $M_V^2$ are separately zero in the gaugeless limit $e\rightarrow 0$, but their ratio is not and proportional to
$\al_t\al_s$. In \figref{fig:vectorbosonself}, we present some sample 
Feynman diagrams which contribute to $\deltatwo M_V^2$. In this
calculation we also set $m_b=0$. It turns out that the contribution of
the left-handed sbottom to the $Z$ boson self-energy is
zero. The first term in \eqref{eq:dvexpression} is proportional to the
correction to the $\rho$ parameter. The QCD correction to this
parameter arising from heavy (s)quark exchange has been computed
in the SM \cite{Djouadi:1987gn,Djouadi:1987di} and MSSM
\cite{Djouadi:1996pa,Djouadi:1998sq}. Our result reproduces the SM
result $\Delta \rho =- (1+ \pi^2/3)\al_s m_t^2/(8 \pi^3 v^2)$, which is
computed within dimensional regularization, while our calculation is
performed in dimensional reduction. The explicit evaluation of the UV divergent
part of  $\deltatwo v/v$ shows that it is related to $\deltatwo Z_{H_u}$ as
\be 
\fr{\deltatwo v}{v}\Big|_{\text{div}} =\fr{s_\beta^2}{2}\deltatwo
Z_{H_u} \,,
\ee
which is to be expected according to \cite{Sperling:2013eva,Sperling:2013xqa}.
\item The ratio of the vacuum expectation values of the Higgs
  doublets, $\tan\!\beta$, is renormalized as a $\DRb$ parameter and
  its counterterm is given by 
\cite{Brignole:1992uf,Chankowski:1992ej,Chankowski:1992er,Dabelstein:1994hb,Dabelstein:1995js,Freitas:2002um} 
\begin{equation}
 \deltatwo\tan\beta=\frac{1}{2}\tan\!\beta\big(\deltatwo
 Z_{H_u}-\deltatwo Z_{H_d}\big)\big|_{\text{div}}=
 \frac{1}{2}\tan\!\beta\,\deltatwo Z_{H_u}\big|_{\text{div}}  \,.
\end{equation}

\item The counterterms of the remaining $\DRb$ parameters 
  $|\lambda|,|\kappa|,v_s,\text{Re}A_{\kappa}$ and $\varphi_y$ are required
  to cancel the UV divergent parts of five independent
  self-energies of the neutral Higgs bosons. As a result, we end up
  with the solution 
 \begin{align}
 \deltatwo|\lambda|&=\frac{-|\lambda|}{2}\big(\deltatwo Z_{H_u}  c_\beta^2+2\fr{\deltatwo v}{v}\Big|_{\text{div}}\big)=\frac{-|\lambda|}{2}\deltatwo Z_{H_u}\,,\\
\deltatwo|\kappa|&=\frac{-|\kappa|}{2}\big(-\deltatwo Z_{H_u} s_\beta^2+2\fr{\deltatwo v}{v}\Big|_{\text{div}}\big)=0\,,\\
\deltatwo v_s&=\frac{-v_s}{2}\big(-\deltatwo Z_{H_u} s_\beta^2+2\fr{\deltatwo v}{v}\Big|_{\text{div}}\big)=0\,,\\
\deltatwo\text{Re}A_{\kappa}&=0\,,\\
\deltatwo \varphi_y&=0\,.
\end{align}
It turns out that only the counterterm of $|\lambda|$ is non-zero. All
other parameters, $|\kappa|,v_s,\text{Re}A_{\kappa}$ and  $\varphi_y$, need not
be renormalized at order ${\cal O}(\al_t\al_s)$.
\end{itemize}

Finally we would like to comment on the cancellation of the
UV-divergences and the differences compared to the respective MSSM
calculation. It is a well known fact that in the calculation of the $\al_t\al_s$  
contributions to the Higgs masses with vanishing external momentum in
the MSSM the counterterm $\deltatwo Z_{H_u}$ is not needed to cancel
the divergences. Furthermore no counterterm for the VEV
renormalization appears. The latter is straight forward to
understand. As can be read off from the counterterm mass matrix as given in
\appen{sec:dMH2loop} all terms including $\deltatwo v$
are proportional to $\lambda$ or $\kappa$  so that they vanish in the
MSSM limit. A more subtle argument for the non-existence of the
$\deltatwo v$ contributions in the MSSM, which can also be applied to the
field renormalization constant, can be made when investigating the
order of the considered corrections. On the one hand in the MSSM the
neutral Higgs self-energies of the doublet-doublet mixing with
vanishing external momenta are proportional to $\al_t\al_s m_t^2$. On
the other hand, in the NMSSM there are also  
mixings between Higgs doublet and singlet components and their
self-energies that go with $\al_t\al_s$  (no additional factors of $m_t$).
This is exactly the order of the $\deltatwo v$ and $\deltatwo Z_{H_u}$
contributions. This is confirmed by the fact, that neglecting
$\deltatwo Z_{H_u}$ and $\deltatwo v$ a UV-finite result can be
obtained for all self-energies except for the one mixing the doublet and
singlet components. Turning on these contributions, however, all
results are UV-finite.

\subsection{Tools and checks}
In two independent calculations we have employed \texttt{FeynArts}
\cite{Kublbeck:1990xc,Hahn:2000kx} for the generation of the
amplitudes using a model file created by \texttt{SARAH} 
\cite{Staub:2009bi,Staub:2010jh,Staub:2012pb,Staub:2013tta}. The
contraction of the Dirac and $\gamma_5$ matrices was done with 
\texttt{FeynCalc} \cite{Mertig:1990an}. The reduction to master
integrals was performed using the program \texttt{TARCER}
\cite{Mertig:1998vk}, which is based on a reduction algorithm proposed
by Tarasov \cite{Tarasov:1996br,Tarasov:1997kx} and which is 
included in \texttt{FeynCalc}.  Additional checks of the calculation
of the self-energy diagrams have been  
carried out applying in-house mathematica routines for the evaluation of scalar
 self-energy diagrams but also using the programs 
{\tt OneCalc} and {\tt TwoCalc} \cite{Weiglein:1993hd, Weiglein:1995qs} 
for the contraction of Dirac matrices, 
the evaluation of Dirac traces and the tensor reduction of the integrals in 
combination with the package  \texttt{FeynArts} for the amplitude generation. 
We have applied dimensional reduction \cite{Siegel:1979wq,Stockinger:2005gx}
 in the manipulation of the Dirac algebra and in the tensor reduction. 
In the MSSM, this has been shown to preserve SUSY at order ${\cal
  O}(\al_t\al_s)$ \cite{Hollik:2005nn}. In the NMSSM there are no
structurally new terms that could violate this, so that SUSY should be
preserved here as well without the necessity to add a SUSY restoring
counterterm. In our calculation no $\gamma_5$ terms appear that
require a special treatment in $D$ dimensions, so that we take $\gamma_5$ to be 
anti-symmetric with all other  Dirac matrices. The results of these 
computations are in full agreement. \s

Furthermore, we compared all  doublet-doublet mixing Higgs
self-energies with the results of the complex
MSSM~\cite{Heinemeyer:2007aq}, setting all possible complex phases
non-zero. It should be noted that, in the MSSM, at tree level, there
exists no physical phase in the Higgs sector, and accordingly, in the
MSSM calculation of Ref.~\cite{Heinemeyer:2007aq} 
the unphysical phases have been rotated away. In order to
compare our NMSSM results with the MSSM results, the phase
of $\mu$ in the MSSM had to be chosen as $\varphi_\mu^{\text{MSSM}} =
\varphi_\lambda + \varphi_s + \varphi_u$. We found perfect agreement, 
provided that $\deltatwo v/v$ was turned off. 
We have also compared with the existing NMSSM results
\cite{Degrassi:2009yq} where all parameters are  real and defined in
the $\DRb$ scheme. Our results are in full agreement with these results as
well.\footnote{Note, however, that in the
  translation from the OS value $v^\OS$ to the $\DRb$ value $v^{\DRb}$
  Ref.~\cite{Degrassi:2009yq} did not include the necessary $\deltatwo v$ term.} 

\section{Numerical Analysis \label{sec:numerical}}
\subsection{Input Parameters and Constraints \label{sect-constraints}}
We have performed a scan in the NMSSM parameter space in order to
find an NMSSM scenario that is in accordance with the experimental
Higgs data. The accordance has been checked by using the programs {\tt
  HiggsBounds} \cite{Bechtle:2008jh,Bechtle:2011sb,Bechtle:2013wla}
and {\tt HiggsSignals} \cite{Bechtle:2013xfa}. 
The program {\tt HiggsBounds} requires as inputs the effective couplings
  of the Higgs bosons of the investigated model, normalized to the
  corresponding SM values, as well as the masses, the widths and the
  branching ratios of the Higgs bosons. This allows then to check for
  the compatibility with the non-observation of the SUSY Higgs bosons,
  in particular whether or not the Higgs spectrum is excluded at the
  95\% confidence level (CL) in view of the LEP, Tevatron and LHC
  measurements. The package {\tt HiggsSignals} uses the same input and
validates the compatibility of the SM-like Higgs boson with the
Higgs observation data. A $p$-value is given, which we demanded to
be at least 0.05, corresponding to a non-exclusion at 95\% CL. For the
computation of the Higgs boson masses, the effective couplings, the decay widths
and branching ratios of the SM and
NMSSM Higgs bosons, the Fortran code {\tt NMSSMCALC} \cite{Baglio:2013iia} is
used. Besides the masses with the newly implemented two-loop
corrections, it provides the SM and NMSSM decay widths and branching
ratios including the state-of-the-art higher order corrections. In
particular, the effective NMSSM Higgs coupling to the gluons
normalized to the corresponding coupling of a SM Higgs boson is
obtained by taking the ratio of the partial width for the Higgs decay
into gluons in the NMSSM and the SM, respectively. The program {\tt
  NMSSMCALC} takes into account the QCD corrections up to
next-to-next-to-next-to leading order in the limit of heavy quark
\cite{Inami:1982xt,Djouadi:1991tka,Spira:1993bb,Spira:1995rr,Kramer:1996iq,Chetyrkin:1997iv,Chetyrkin:1997un,Schroder:2005hy,Chetyrkin:2005ia,Baikov:2006ch}
and squark \cite{Dawson:1996xz,Djouadi:1996pb} masses. They can be
taken over from the SM, respectively, MSSM case. As the electroweak corrections are
unknown for SUSY Higgs decays, they are consistently neglected also in
the SM decay width. In the same way we proceed for the 
loop-mediated effective Higgs coupling to the photons. In this case
the next-to-leading order QCD corrections to quark and squark loops
including the full mass dependence for the quarks
\cite{Spira:1995rr,Zheng:1990qa,Djouadi:1990aj,Dawson:1992cy,Djouadi:1993ji,Melnikov:1993tj,Inoue:1994jq}
and squarks \cite{Muhlleitner:2006wx} are taken into 
account. Again electroweak corrections, which are not known for the
SUSY case, are neglected also in the SM. \s

The parameter point fulfilling the above constraints and that we use
in our numerical analysis is given by the following input
parameters. The SM parameters \cite{Agashe:2014kda,Jegerlehner:2011mw}
are 
\begin{align}
\alpha(M_Z)&=1/128.962 \,, &\alpha^{\overline{\mbox{MS}}}_s(M_Z)&=
0.1184 \,, &M_Z&=91.1876\, 
\gev\,, \\ \non
M_W&=80.385\,\gev \,,   &m_t&=173.5\,\gev \,,
&m^{\MSb}_b(m_b^{\MSb})&=4.19\,\gev \;. \label{eq:param1}
\end{align} 
The running strong coupling constant $\al_s$ is evaluated by using the
SM renormalization group equations at two-loop order. The light quark
masses, which have only a small influence on the loop results, are
chosen as 
\beq
m_u=2.5\,\mev\; , \quad m_d=4.95\,\mev\; , \quad m_s=101\, \mev \quad
\mbox{and} \quad m_c=1.27\, \gev \;. \label{eq:param2}
\eeq
\noindent 
The soft SUSY breaking masses and trilinear couplings 
have been set to 
\beq
&&  m_{\tilde{u}_R,\tilde{c}_R} = 
m_{\tilde{d}_R,\tilde{s}_R} =
m_{\tilde{Q}_{1,2}}= m_{\tilde L_{1,2}} =m_{\tilde e_R,\tilde{\mu}_R} = 3\;\mbox{TeV}\, , \;  
m_{\tilde{t}_R}=1170\,\gev \,,\; \non \\ \non
&&  m_{\tilde{Q}_3}=1336\,\gev\,,\; m_{\tilde{b}_R}=1029\,\gev\,,\; 
m_{\tilde{L}_3}=2465\,\gev\,,\; m_{\tilde{\tau}_R}=300.5\,\gev\,,
 \\ 
&& |A_{u,c,t}| = 1824\,\gev\, ,\; |A_{d,s,b}|=1539\,\gev\,,\; |A_{e,\mu,\tau}| = 1503\,\gev\,,\; \\ \non
&& |M_1| = 862.3\,\gev,\; |M_2|= 201.5\,\gev\,,\; |M_3|=2285\,\gev \,,\\ \non
&&  \varphi_{A_{d,s,b}}=\varphi_{A_{e,\mu,\tau}}=\pi\,,\; 
\varphi_{A_{u,c,t}}=\varphi_{M_1}=\varphi_{M_2}=\varphi_{M_3}=0
 \;. \label{eq:param4}
\eeq
For the remaining input parameters we chose
\beq
&& |\lambda| = 0.629 \;, \quad |\kappa| = 0.208 \; , \quad |A_\kappa| = 179.7\,\gev\;,\quad 
|\mu_{\text{eff}}| = 173.7\,\gev \;, \non \\ 
&&\varphi_{\lambda}=\varphi_{\mu_{\text{eff}}}=\varphi_u=0\;, \quad \varphi_{\kappa}=\pi \;, 
\quad \tan\beta = 4.02 \;,\quad M_{H^\pm} = 788 \,\gev \;.
\eeq
In compliance with the SLHA, we take $\mu_{\text{eff}}$ as input
parameter, from which $v_s$ and $\varphi_s$ can be obtained through
Eq.~(\ref{eq:mueff}). Note that the parameters $\lambda, \kappa,
A_\kappa, \mu_{\text{eff}}, \tan\beta$ as well as the soft 
SUSY breaking masses and trilinear couplings are understood as $\DRb$
parameters at the scale $\mu_R = M_s$\footnote{For $\tan\beta$ this is
only true, if it is read in from the block EXTPAR as done in {\tt
  NMSSMCALC}. Otherwise it is the 
$\DRb$ parameter at the scale $M_Z$.}, while the charged Higgs mass is
an OS parameter. The SUSY scale $M_s$ is set to be 
\beq
M_s = \sqrt{m_{\tilde Q_3}m_{\tilde t_R}} \;.
\eeq
Our chosen parameter values guarantee the supersymmetric particle
spectrum to be in accordance with present LHC searches for SUSY particles 
\cite{Aad:2012tx,Aad:2012yr,Aad:2013ija,Aad:2014qaa,Aad:2014wea,Aad:2014bva,Aad:2014kra,Chatrchyan:2013xna,CMSPAS13008,Chatrchyan:2013fea,Chatrchyan:2013mya,CMSPAS13018,CMSPAS13019,Khachatryan:2014doa,CMSPAS14011}. In the following we will drop the subscript
'$\text{eff}$' for $\mu$. Furthermore, we will use the expressions OS
and $\DRb$ in order to refer to the renormalization in the top/stop
sector. \s

\subsection{Results  \label{sec:results}}
The masses that we obtain for the chosen scenario at tree level,
at one-loop and at two-loop order when using the OS scheme in the
top/stop sector are shown in Tab.~\ref{tab:massOS}. The results for
the $\DRb$ scheme in the top/stop sector can be found in
Tab.~\ref{tab:massDR}. The tables also show the main singlet/doublet
and scalar/pseudoscalar component of the respective mass eigenstate. 
For completeness we furthermore give the values of the tree-level stop
masses obtained for the $\DRb$ and for the OS scheme, respectively,
\beq
\begin{array}{llll}
\DRb &:& m_{\tilde{t}_1} = 1126 \mbox{ GeV} \;, \qquad & m_{\tilde{t}_2} = 1387
\mbox{ GeV} \;, \\
\mbox{OS} &:& m_{\tilde{t}_1} = 1144 \mbox{ GeV} \;, \qquad &
m_{\tilde{t}_2} = 1421 \mbox{ GeV} \;.
\end{array}
\eeq
The $\DRb$ top mass in our scenario is given by $m_t^{\DRb} =
143.14$~GeV. \s

\begin{table}[t]
\begin{center}
 \begin{tabular}{|l||c|c|c|c|c|}
\hline
 &${H_1}$&${H_2}$&${H_3}$&${H_4}$&${H_5}$\\ \hline \hline
mass tree [GeV] &79.15&103.55&146.78&796.62&803.86\\
main component&$h_s$&$h_u$&$a_s$&$h_d$&$a$\\ \hline
mass one-loop [GeV] &103.45&129.15&139.84&796.53&802.94\\
main component&$h_s$&$a_s$&$h_u$&$h_d$&$a$\\ \hline
mass two-loop [GeV] &103.00&126.20&128.93&796.45&803.07\\
main component&$h_s$&$h_u$&$a_s$&$h_d$&$a$\\ \hline
\end{tabular}
\caption{Masses and main components of the neutral Higgs bosons at tree, one- and two-loop level as obtained using OS renormalization in the top/stop sector.}
\label{tab:massOS}
\end{center}
\end{table}
\begin{table}[b]
\begin{center}
 \begin{tabular}{|l||c|c|c|c|c|}
\hline
 &${H_1}$&${H_2}$&${H_3}$&${H_4}$&${H_5}$\\ \hline \hline
mass tree [GeV] &79.15&103.55&146.78&796.62&803.86\\
main component&$h_s$&$h_u$&$a_s$&$h_d$&$a$\\ \hline
mass one-loop [GeV] &102.80&120.52&128.80&796.36&803.09\\
main component&$h_s$&$h_u$&$a_s$&$h_d$&$a$\\ \hline
mass two-loop [GeV] &103.09&124.52&128.91&796.36&803.03\\
main component&$h_s$&$h_u$&$a_s$&$h_d$&$a$\\ \hline
\end{tabular}
\caption{Masses and main components of the neutral Higgs bosons at tree, one- and two-loop level as obtained using $\drbar$ renormalization in the top/stop sector.}
\label{tab:massDR}
\end{center}
\end{table}
The scenario features three light Higgs bosons that are rather close
in mass. For a meaningful interpretation of the results for the mass
corrections, the Higgs bosons with a similar admixture have
to be compared and not the ones 
corresponding to each other due to their mass ordering. Thus $H_2$ is
$h_u$ dominated at tree level, however in the OS scheme at one-loop
level this role is taken over by $H_3$, so that these two states have to be
compared. Hence in the following plots we will label the Higgs bosons
not by their mass ordering but according to their main
components. Note that in our scenario the $h_u$ dominated Higgs boson
is the SM-like Higgs boson. \s

Since the lightest Higgs boson $H_1$, which is scalar-like singlet dominated, has a
small tree-level mass value,  the one-loop corrections are
rather large as expected. The two-loop corrections for $h_s$ are below
1\%. At tree level the $h_u$-dominated Higgs
boson is $H_2$. With the main loop contributions stemming from the
top/stop sector the $h_u$-type Higgs boson hence receives important one-loop
corrections of ${\cal O}(16\%)$ in the $\DRb$, respectively ${\cal
  O}(35\%)$ in the OS scheme. Adding the two-loop corrections reduces
the mass value by $\sim 10\%$ in the OS scheme and increases it by
$\sim 3\%$ in the $\DRb$ scheme, so that finally the two-loop masses
differ by about 1.3\% in the two renormalization schemes. For the
singlet-dominated pseudoscalar-like, {\it i.e.}~$a_s$-like Higgs boson
the one-loop corrections in both schemes are at the 10\% level and below
1\% at two-loop order. The heavy Higgs bosons $H_4$ and $H_5$ finally
with masses around 800 GeV are hardly affected by loop corrections. \s

\begin{figure}[t]
 \includegraphics[width=0.5\textwidth]{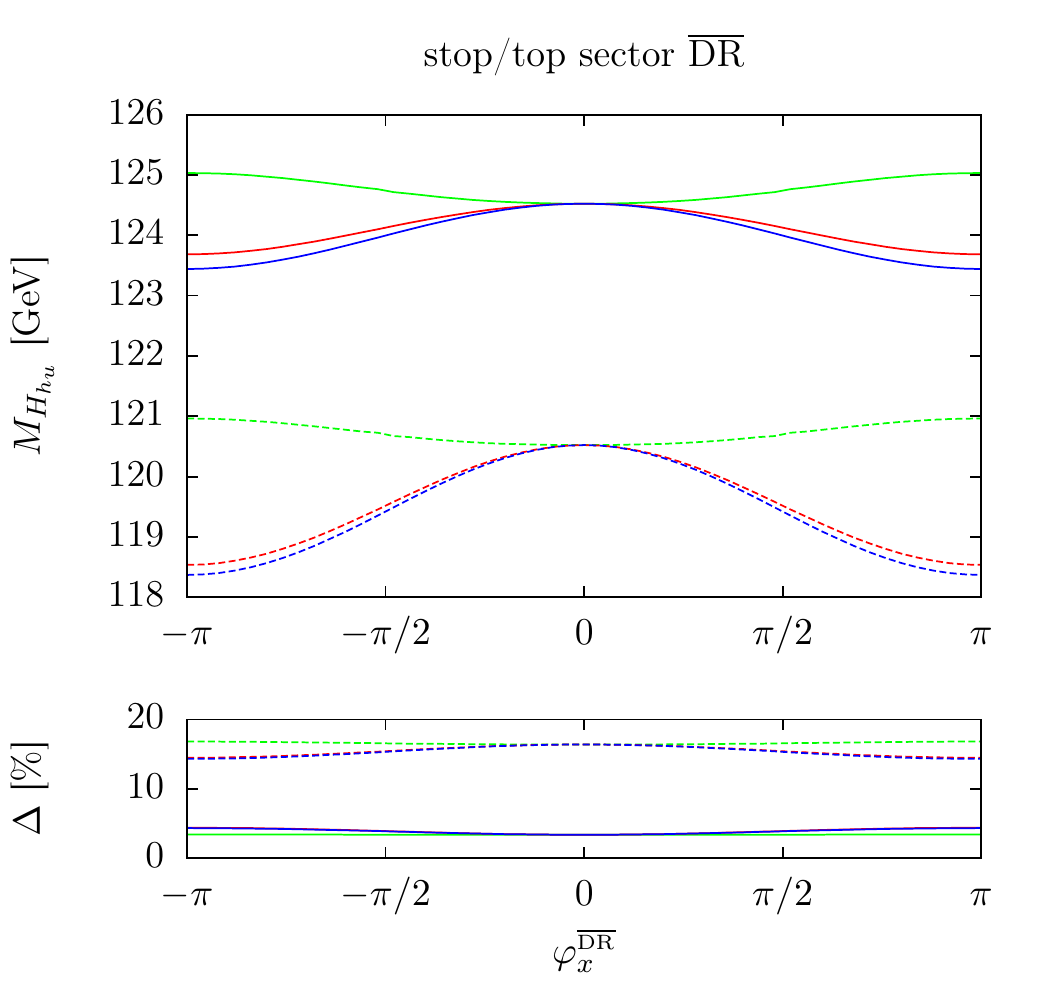} \includegraphics[width=0.5\textwidth]{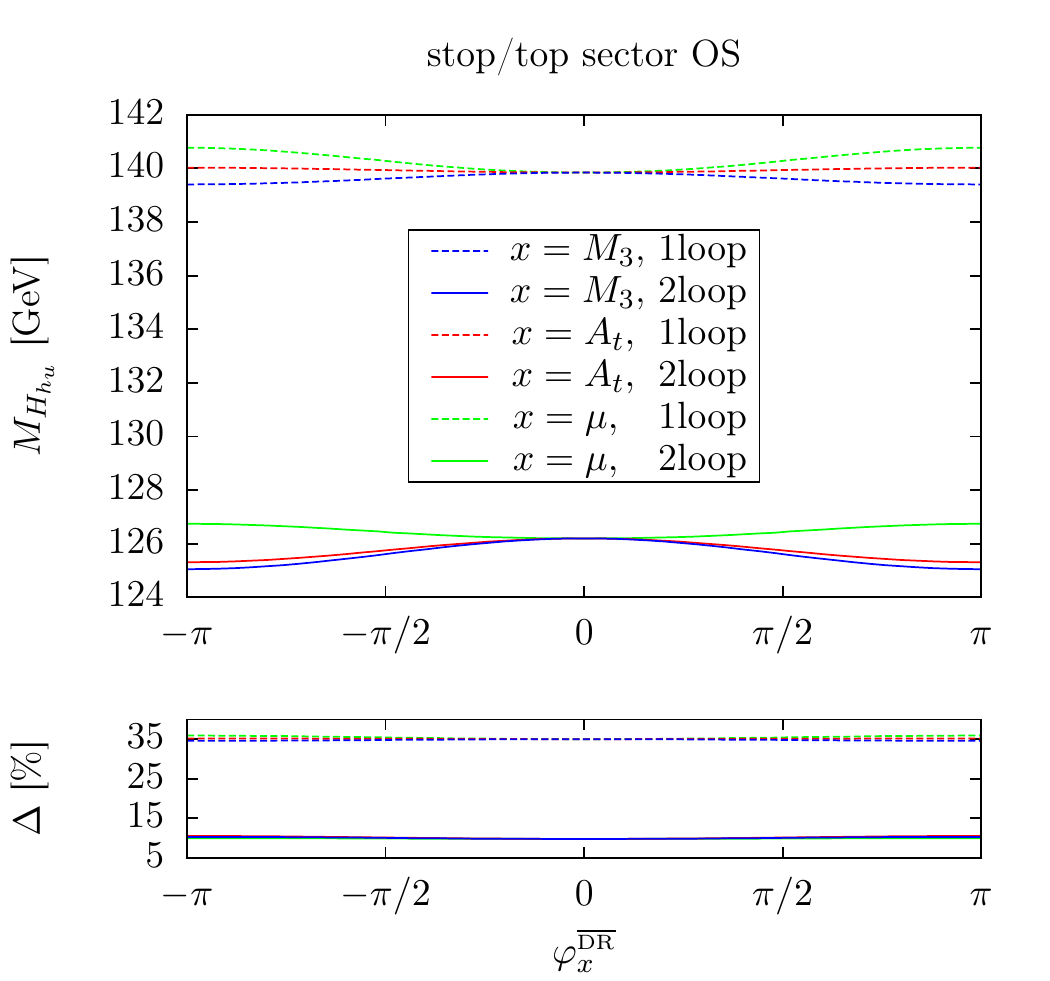}
\caption{Upper Panels: One-loop (dashed line) and two-loop (solid line)
  mass of the SM-like Higgs boson as a function of the phases
  $\varphi_{\mu}$ (green/grey), $\varphi_{A_t}$ (red/black upper) and $\varphi_{M_3}$
  (blue/black lower). Lower Panels: Size of the relative correction of $n^{th}$
  order to the mass of the SM-like Higgs boson with respect to the
  $(n-1)^{st}$ order -- {\it
    i.e.}~$\Delta=|M_{H_{h_u}}^{(n)}-M_{H_{h_u}}^{(n-1)}|/M_{H_{h_u}}^{(n-1)}$ -- in 
  percent as a function of the phases $\varphi_{\mu}$ (green/grey),
  $\varphi_{A_t}$ (red) and $\varphi_{M_3}$ (blue) for $n=2$ (solid
  line) and $n=1$ (dashed line). On the left-hand side $\drbar$ renormalization
  was employed in the top/stop sector and OS renormalization on the right-hand
  side.}
\label{fig:phasevariation}
\end{figure}
Figure~\ref{fig:phasevariation} shows the dependence of the one- and
two-loop corrections to the mass of the $h_u$-like Higgs boson on the
phases $\varphi_{A_t}$, $\varphi_{M_3}$ and  $\varphi_{\mu}$ for
the $\DRb$ renormalization scheme as well as for the OS scheme in the
top/stop sector.\footnote{Note that we vary the phases here for
  illustrative purposes also up to values that may already be excluded 
  by the experiments.} We start from the above defined parameter point and
turn on separately one of the three phases. The corrections are
displayed only for the $h_u$-like Higgs boson, because it is affected
the strongest by the ${\cal O} (\alpha_t\alpha_s)$ corrections. For
both schemes the phase 
dependence displayed at two-loop level is very similar. For the here investigated 
scenario the strongest dependence occurs for the variation of the
phase of $M_3$. The dependence on $\varphi_{A_t}$ is slightly less
pronounced, but comparable, whereas the curve for $\varphi_{\mu}$ is
significantly flatter. We have taken care to vary $\varphi_{\mu}$ in such a way that
the CP-violating phase, which appears already at tree level in the Higgs sector,
{\it i.e.}~$\varphi_y = \varphi_{\kappa} -\varphi_{\lambda} + 2\varphi_{s}-\varphi_{u}$, 
remains at zero. This implies that $\varphi_{\lambda}$ and $\varphi_{s}$ were varied
at the same time, in particular $\varphi_\lambda= 2\varphi_s= 2/3
\varphi_\mu$. The phases $\varphi_\kappa$ and
$\varphi_u$ are kept zero. The correlation, respectively,
anticorrelation of the dependences of the loop corrections on the
various phases can be traced back to the observation, that the
influence of the phases $\varphi_{M_3}$, $\varphi_{A_t}$ and
$\varphi_\mu$ can be described by two independent phase combinations
$\varphi_1$ and $\varphi_2$ given by 
\beq
\varphi_1 = \varphi_\mu + \varphi_{A_t} \qquad \mbox{ and } \qquad
\varphi_2 = \varphi_{M_3} - \varphi_{A_t} \;.
\eeq 
The relative influence of $\varphi_{1,2}$ on the loop corrections can
then explain the observed behaviour. \s 

At the one-loop level the results for the two different
renormalization schemes in the top/stop sector seem quite different at
first sight. However, it has to be kept in mind, that in the $\DRb$
scheme the OS input value for the top mass has to be converted to the
$\DRb$ top mass and while doing so the finite counterterm to the top
mass, which in the OS scheme is included in the two-loop calculation, is
already induced at one-loop level in the value of the $\DRb$
mass. Therefore some corrections of order 
${\cal O}(\alpha_t\alpha_s)$, which in the OS scheme only appear at the two-loop
level, are moved to the one-loop level. This is also the reason why the
loop-corrected masses in the $\DRb$ scheme show a dependence on the phase
$\varphi_{M_3}$ already at the one-loop level, although genuine
diagrammatic gluino corrections only appear at two-loop level. For the
OS scheme this dependence at one-loop level is due to the
conversion of $A_t$ and of the soft SUSY breaking masses, which in the
SLHA input are $\DRb$ parameters, to the OS scheme. The lower panels
of Fig.~\ref{fig:phasevariation} display the relative loop corrections
of $n$-loop order compared to the one at $(n-1)$-loop order ($n=1,2$),
\beq
\Delta=\frac{|M_{H_{h_u}}^{(n)}-M_{H_{h_u}}^{(n-1)}|}{M_{H_{h_u}}^{(n-1)}} \;.
\eeq
As can be read off from the plots, the two-loop corrections relative to the one-loop
mass are of course  
smaller than the one-loop corrections relative to the tree-level mass,
which amount to about 15\% in the $\DRb$ scheme and to about 35\% when
adopting OS renormalization. Still the two-loop corrections amount to
some $5-10\%$. (In the left lower panel, the lines for the $\varphi_{M_3}$ and the
$\varphi_{A_t}$ dependence lie on top of each other, whereas in the
right lower panel all lines lie nearly on top of each other at the
respective loop order.) \s

\begin{figure}[t]
  \includegraphics[width=0.5\textwidth]{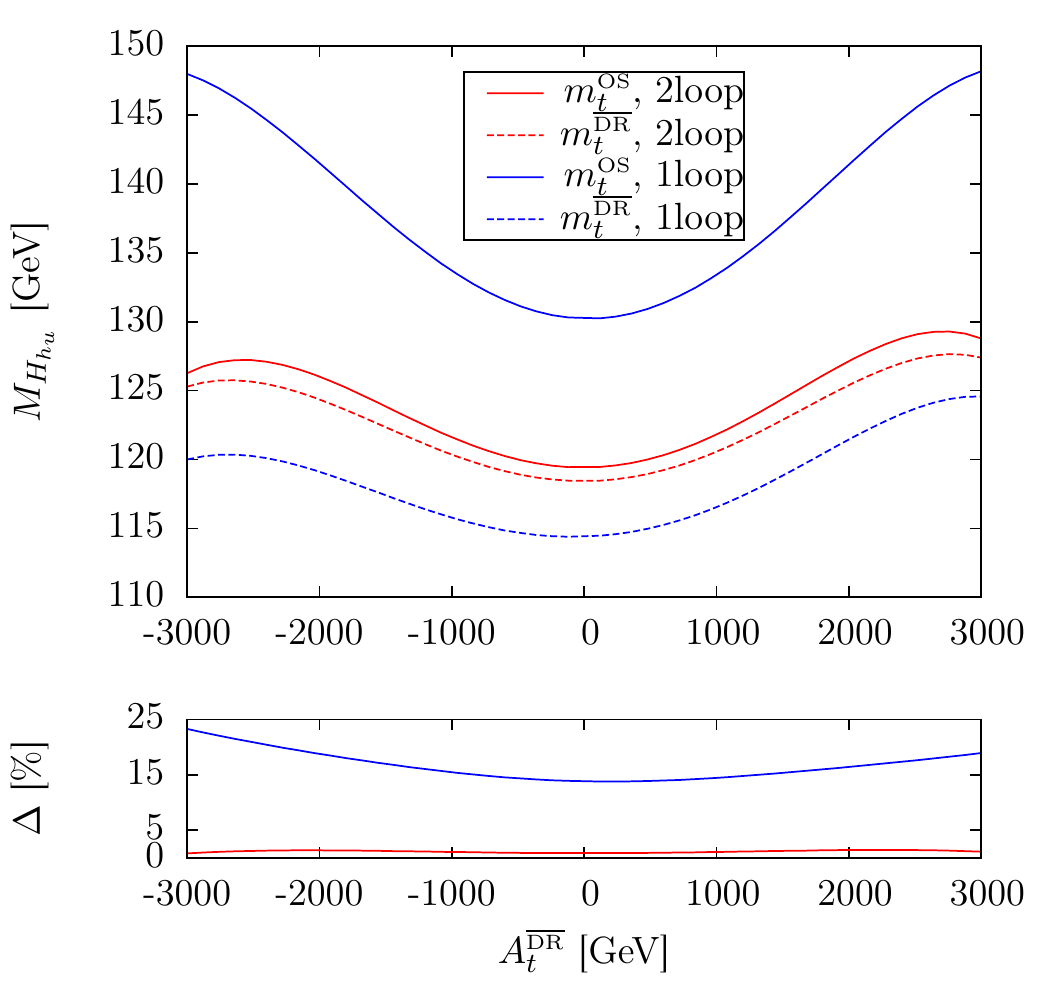}\includegraphics[width=0.5\textwidth]{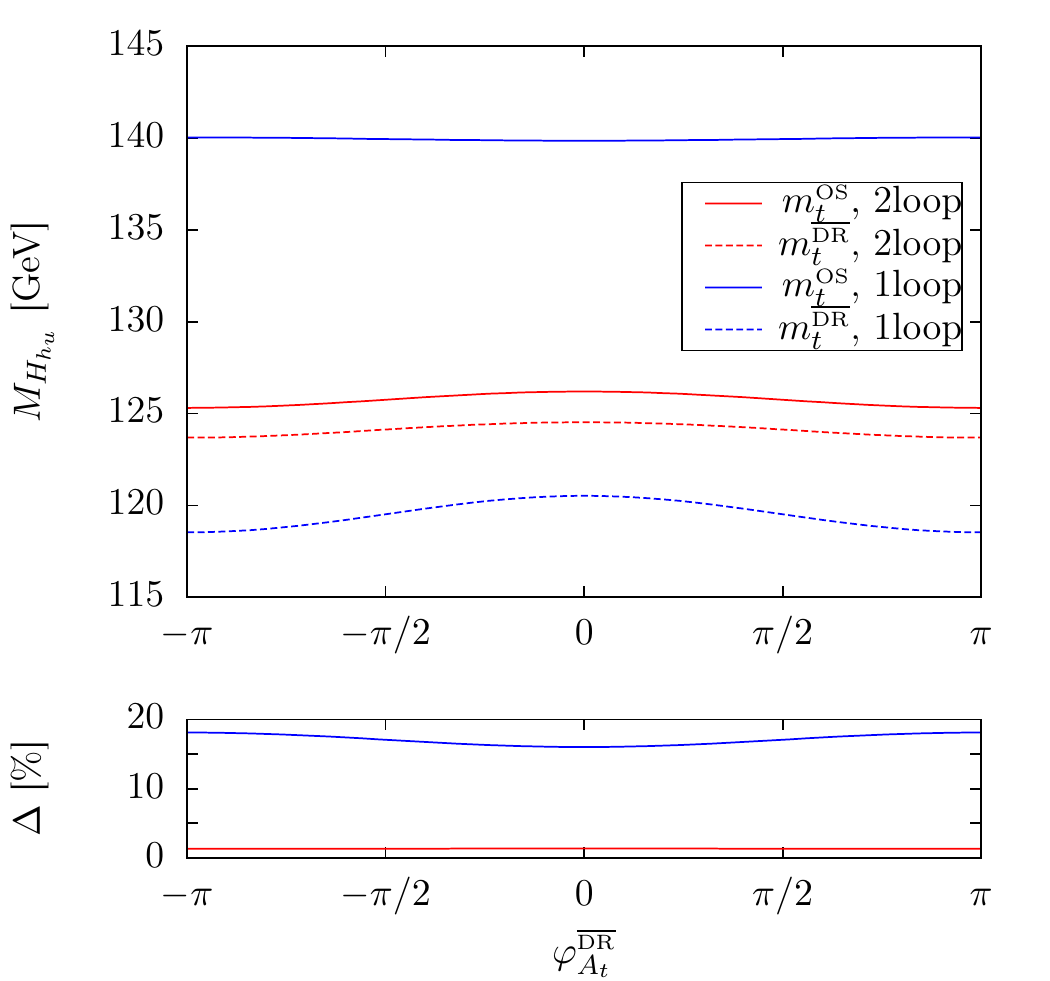}
\caption{Upper Panel: Mass of the $h_u$-like Higgs boson as a function
  of $A_t^{\DRb}$ (left) and $\varphi_{A_t}^{\DRb}$ (right) including
  only one loop corrections (blue/two outer lines) and including also two loop
  corrections (red/two middle lines). For the renormalization of the top and stop
  sector either an OS scheme (solid line) or a $\drbar$ scheme (dashed
  line) is applied. Lower Panel: Absolute value of the relative
  deviation of the result using OS renormalization in the top and stop
  sector with respect to the result using a $\drbar$ scheme --
  {\it i.e.} $\Delta=|M_{H_{h_u}}^{m_t(\drbar)}-M_{H_{h_u}}^{m_t({\tiny
      \mbox{OS}})}|/M_{H_{h_u}}^{m_t({\tiny \drbar})}$ -- in percent as a
  function of $A_t^{\DRb}$ (left) and $\varphi_{A_t}^{\DRb}$ at two
  (red/lower line) and one loop order (blue/upper line).}
\label{fig:onelooptwoloop}
\end{figure}
To provide a rough estimate of the theoretical error due to missing higher
order corrections, Fig.~\ref{fig:onelooptwoloop} shows the one-loop and
two-loop mass of the $h_u$-like Higgs boson as a function of the
$\DRb$ parameters $A_t$ (left) and $\varphi_{A_t}$ (right) for both $\DRb$ and OS 
renormalization in the top/stop sector. The difference between the two
schemes is more pronounced for large absolute values of $A_t$. As
expected the difference in the masses obtained using the two 
schemes,
\beq
\Delta=\frac{|M_{H_{h_u}}^{m_t(\drbar)}-M_{H_{h_u}}^{m_t({\tiny
      \mbox{OS}})}|}{M_{H_{h_u}}^{m_t({\tiny \drbar})}} \;,
\eeq 
becomes much smaller when going from one to two loops. The
lower panels of Fig.~\ref{fig:onelooptwoloop} show that it 
drops from some $15-25\%$ difference to a value below $1.5\%$. This is
an indicator that the theoretical error is also reduced. The
convergence in the $\DRb$ scheme is better than in the OS
scheme. \s 

Note, that the one-loop corrections in the OS top mass scheme are
symmetric with respect to a change of $A_t$, while this is not the
case for the $\DRb$ scheme. This is due to the threshold effects in
the conversion of the top OS to $\DRb$ 
mass. They depend on the sign of $A_t$. In the right plot, the variation
of the loop-corrected masses with $\varphi_{A_t}$ is due to two
effects, the genuine dependence on the phase and the change of the
stop mass values with the phase, where the latter is the dominant
effect. \s 

\begin{figure}[t]
\begin{center}
\includegraphics[width=0.5\textwidth]{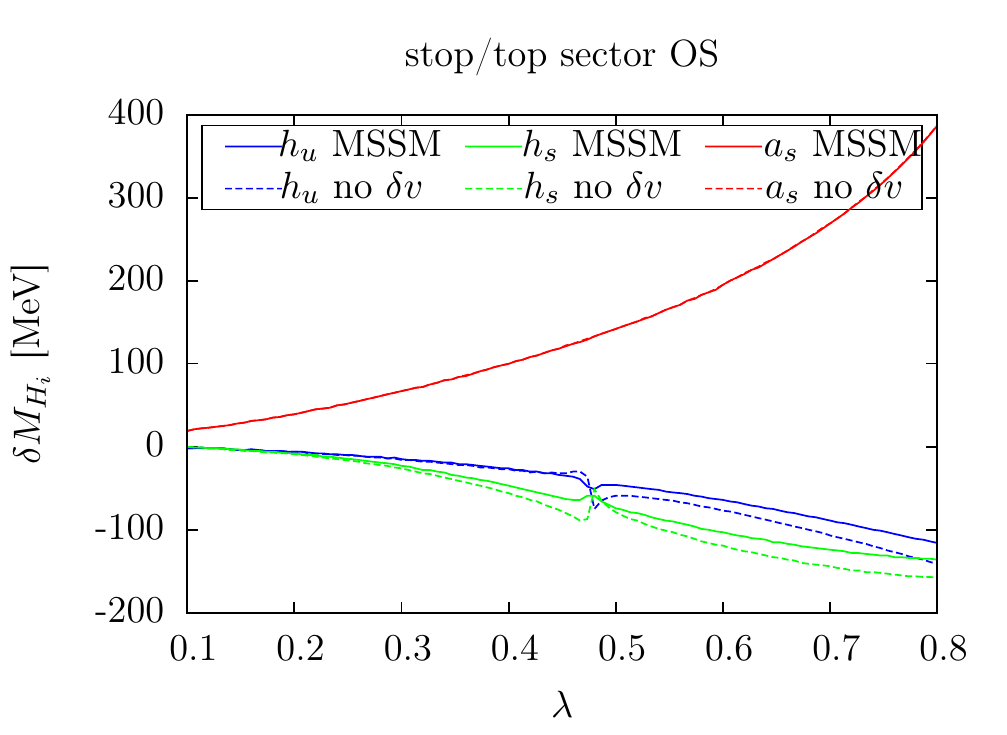}\includegraphics[width=0.5\textwidth]{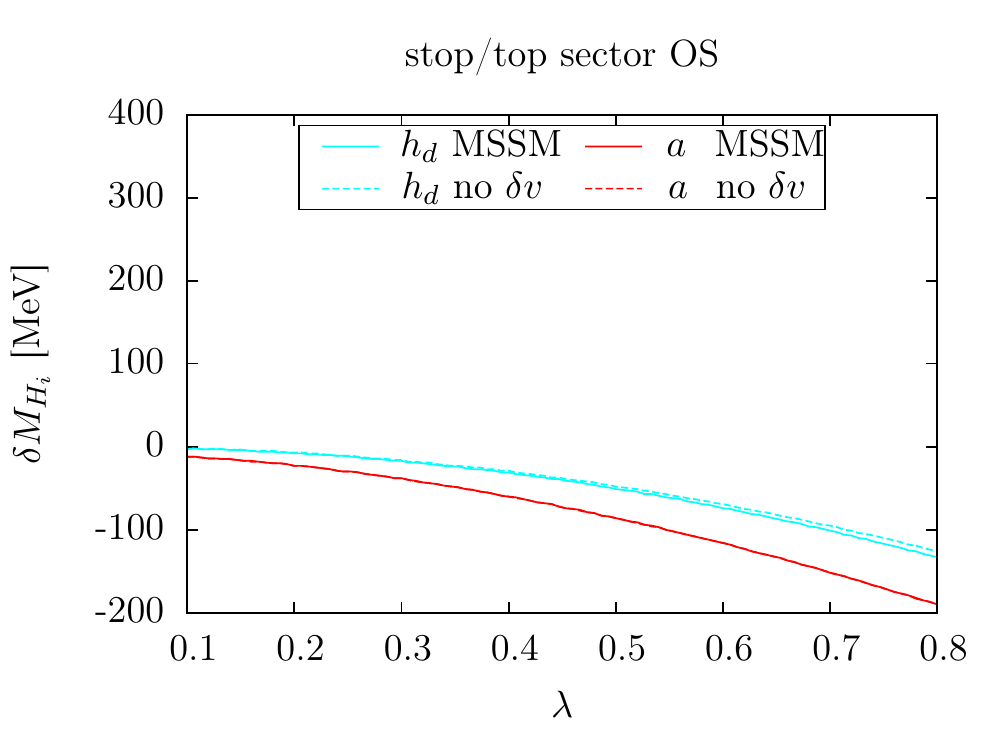} \\
\includegraphics[width=0.5\textwidth]{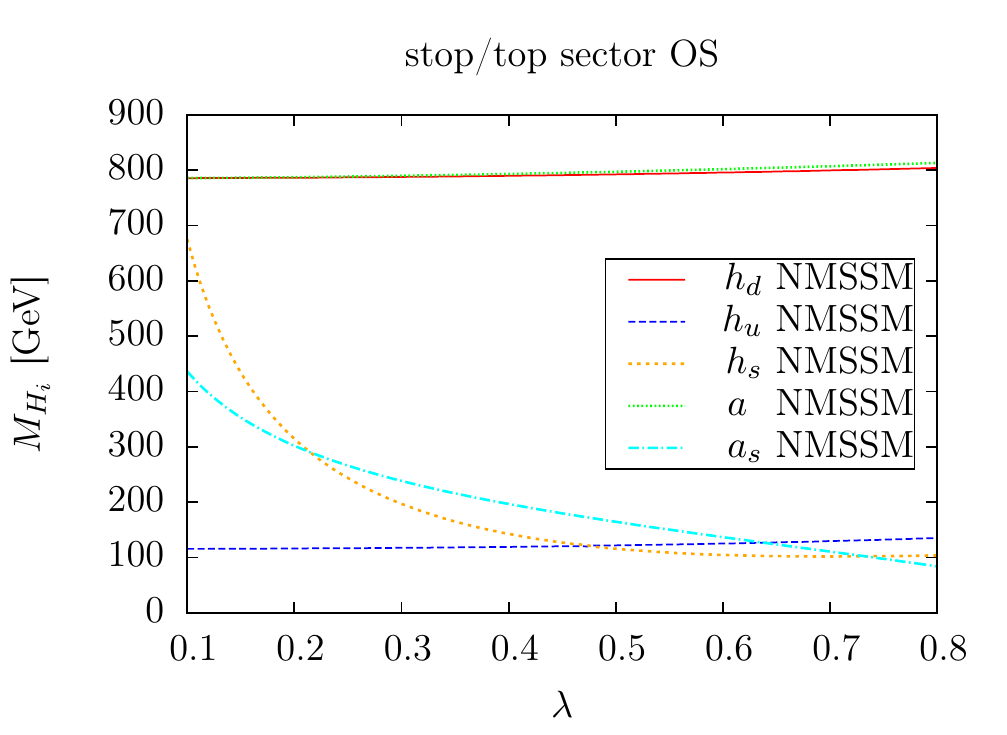}
\end{center}
\caption{Upper plots: The solid lines show the absolute difference of the
  Higgs boson masses obtained when using only the MSSM-like ${\cal O}(\alpha_t
  \alpha_s)$ corrections to the masses and when including the NMSSM
  specific $\alpha_t \alpha_s$; the dotted lines show the absolute
  difference of the masses obtained when neglecting the finite part of
  the two-loop $\deltatwo v/v$ term  and when performing the full NMSSM
  ${\cal O}(\alpha_t\alpha_s)$ calculation. Left: for
  the $h_u$ (blue/black lower lines), $h_s$ (green/grey lower lines)
  and $a_s$ (red/black upper lines) dominated states. Right: for the $h_d$
  (light blue/grey lines) and $a$ (red/black lines) dominated heavy
  Higgs bosons. Lower: 
  Higgs boson masses for the complete ${\cal O}(\alpha_t\alpha_s)$
  NMSSM corrections. All plots as a function of $\lambda$.}  
\label{fig:approximation}
\end{figure}
In Figure~\ref{fig:approximation} (upper part) we illustrate the
impact of the $\deltatwo 
v/v$ contribution, that only in the NMSSM contributes to the Higgs
boson masses at order ${\cal O}(\alpha_t \alpha_s)$, and of the
genuine contributions from the singlet-doublet mixing. In particular
this means that in the two-loop corrections to the Higgs boson masses
we have turned off the finite part of the $\deltatwo v/v$ contribution
in the approximation labeled 'no $\delta v$', and in the
approximation 'MSSM' we have taken the MSSM limit for the two-loop
corrections as specified at the
end of subsection~\ref{sec:higgssector}. As renormalization scheme we
have chosen the OS scheme here. The plots show the
absolute difference in the two-loop corrected mass values for both
approximations as a function of $\lambda$. For illustrative purposes
we allow to vary $\lambda$ here beyond the perturbativity limit, which
is roughly given by $\sqrt{\lambda^2+\kappa^2} <0.7$. While the overall effect is
small and below 1~GeV, it can easily be seen that the importance of the
neglected contributions rises with $\lambda$, as expected. For small
values of $\lambda$ the masses are very close to those obtained when
using only the MSSM two-loop corrections. Regarding the impact of the
finite part of the $\deltatwo v/v$ term it is interesting to note that
neglecting it in the two-loop counterterm mass matrix leads to nearly
the same result (the lines lie on top of each other) for the
pseudoscalar masses as obtained in the MSSM 
limit of the two-loop corrections, where this term vanishes
anyway. Another interesting observation is that it is also possible to
be further away from the full result when neglecting the $\deltatwo
v/v$ contribution than when simply using the MSSM contributions. This
is the case for the Higgs bosons that are dominated by the up-type or
by the singlet component, {\it i.e.}~for the $h_u$ and $h_s$-like Higgs
bosons.\footnote{The small peaks appearing in
  Fig.~\ref{fig:approximation} are due to the fact that here a
  cross-over of the masses of the $h_u$- and $h_s$-like Higgs boson
  occurs.} The lower plot in
Fig.~\ref{fig:approximation} displays the values of the two-loop
corrected Higgs boson masses. The lines for the heavy $a$ and $h_d$
dominated masses lie on top of each other. The plot also shows the
cross-over of the $h_u$ and $h_s$-like Higgs boson masses at $\lambda
\approx 0.475$. There is another cross-over with the
$a_s$-like Higgs boson. However, as we set all phases to zero in this
plot, so that there is no CP mixing, this does not affect the CP-even
light Higgs masses. 
\s

\begin{figure}[th]
\includegraphics[width=0.5\textwidth]{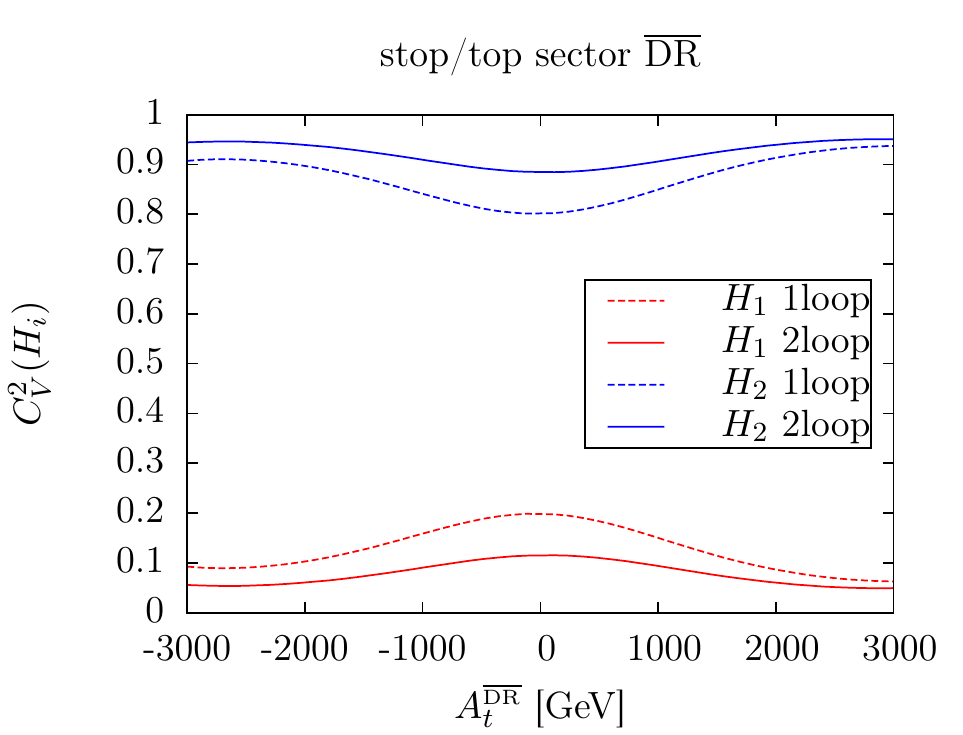}\includegraphics[width=0.5\textwidth]{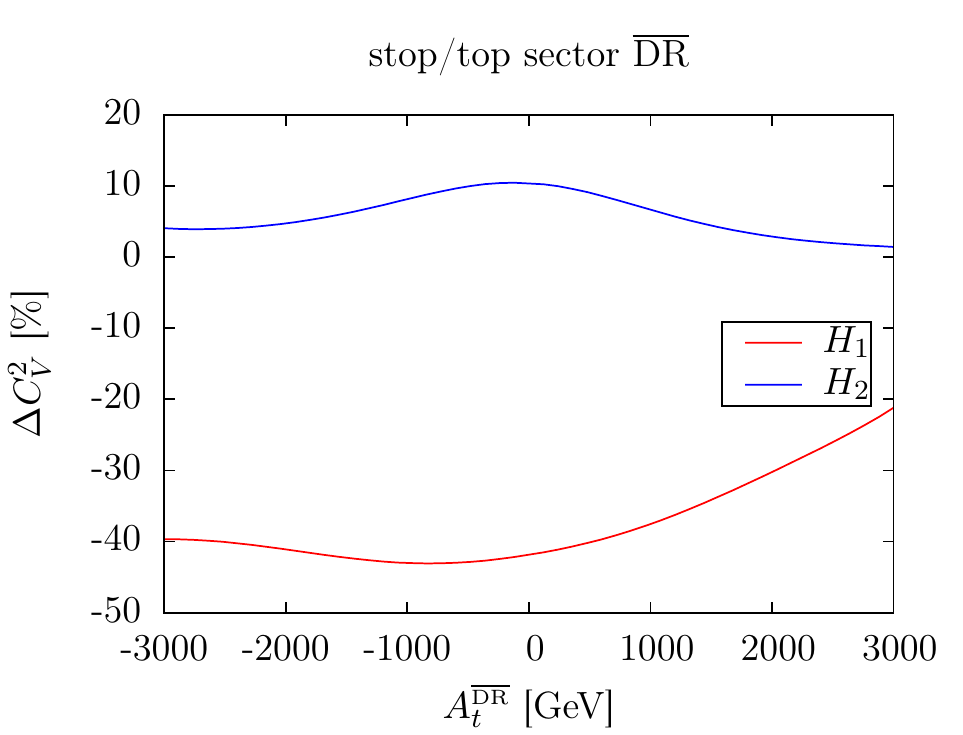}
\caption{Left: Square of the coupling to vector bosons normalized to
  the respective SM coupling for the lightest (red/lower) and
  next-to-lightest Higgs boson (blue/upper) at one-loop order (dashed)
  and at two-loop order (solid). Right: Size of the two-loop
  correction to $C_V^2(H_i)$ relative to the one-loop result for $H_1$
  (red/lower) and $H_2$ (blue/upper); {\it i.e.}~$\Delta
  C_V^2=[(C_V^2)^{\text{(2loop)}}-(C_V^2)^{\text{(1loop)}}]/(C_V^2)^{\text{(1loop)}}$.} 
\label{fig:cvcoup}
\end{figure}

Another interesting question is how the mixing matrix elements are
affected by non-vanishing complex phases. The matrix elements enter
the Higgs couplings and hence influence the Higgs phenomenology. In
general the mixing is hardly influenced by the phases, unless two
of the Higgs bosons are almost mass degenerate and hence share their
various doublet/singlet scalar/pseu\-do\-sca\-lar contributions, as we
explicitly verified. As in this case, however, it turns out that the
mixing elements in the $p^2=0$ approximation differ substantially from
the results obtained from the iterative procedure, also the mass
values need to be obtained in the $p^2=0$ approximation to allow for a
consistent interpretation of the mass values and their related mixing
elements. \s

\begin{figure}[t]
\includegraphics[width=0.5\textwidth]{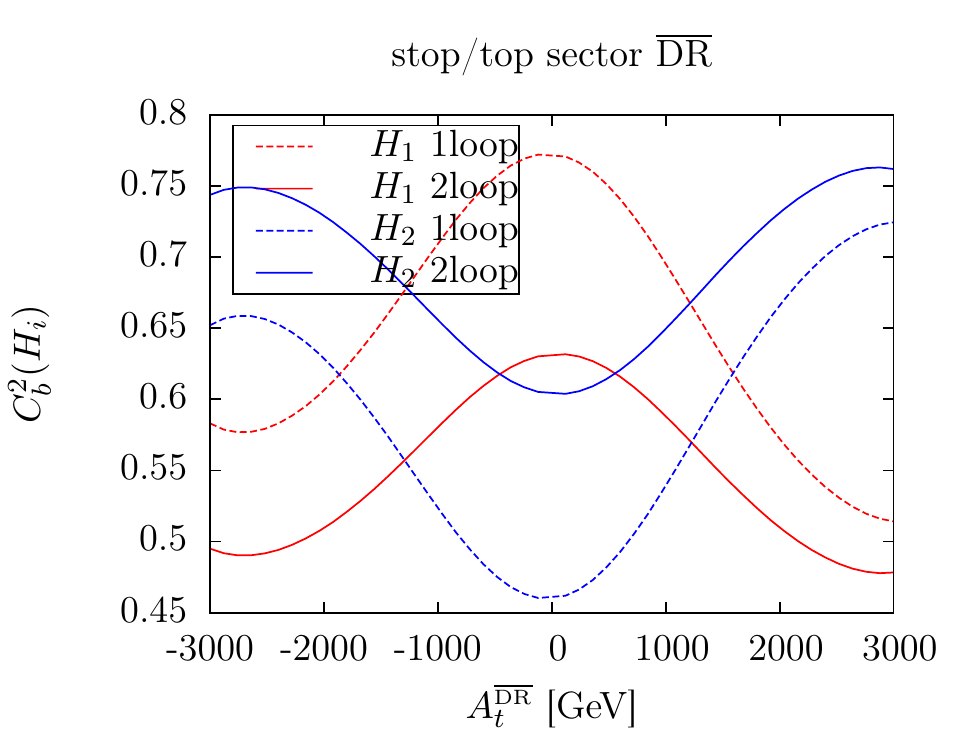}\includegraphics[width=0.5\textwidth]{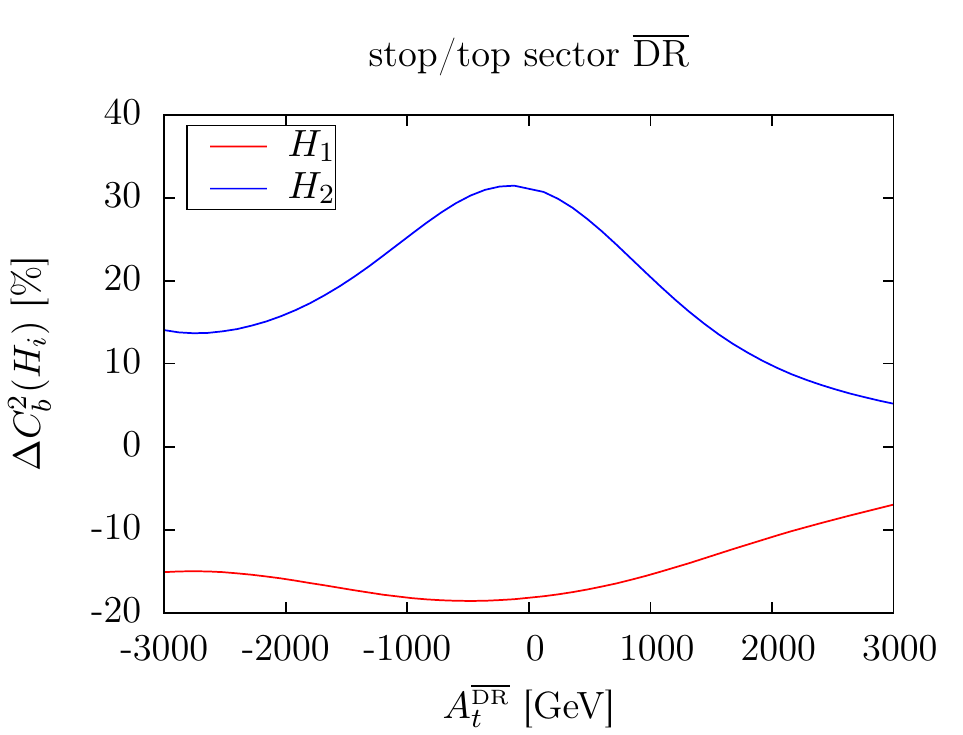}
\caption{Same as Fig.~\ref{fig:cvcoup}, but for the coupling to bottom
  quarks.}
\label{fig:cbcoup}
\end{figure}

In Figs.~\ref{fig:cvcoup} and \ref{fig:cbcoup} we show the influence
of the loop corrections on the couplings $C_V$ to the vector bosons
$V=W,Z$ and $C_b$ to the bottom quarks of the two lightest Higgs
bosons, $H_1$ and $H_2$. The couplings are normalized to the corresponding SM
couplings, so that $C_V$ reads ($i=1,2$)
\beq
C_V (H_i) = {\cal R}^l_{i1} \cos\beta + {\cal R}^l_{i2} \sin\beta \;, 
\eeq
where ${\cal R}^l_{ij}$ denote the matrix elements of the
loop-corrected mixing matrix 
evaluated at zero external momentum, which at tree level has
been defined in Eq.~(\ref{eq:rotationtreelevel}).  
The CP-even Higgs couplings to the bottom quarks are given by 
\beq
C_b (H_i) = \frac{{\cal R}^l_{i1}}{\cos\beta} \;.
\eeq

In Fig.~\ref{fig:cvcoup} (left) the couplings of $H_1$ and $H_2$ to
the vector bosons are displayed at one- and two-loop level as a
function of $A_t^{\drbar}$. In the scenario considered here these are
the only two Higgs bosons which couple non-negligibly to vector
bosons. Since the scenario features a relatively small $\tan\beta$
value of $\sim 4$, the Higgs boson with the largest $h_u$ component  
and hence a sizeable ${\cal R}_{i2}^l$ has the largest coupling to
vector bosons. Therefore the coupling of 
$H_2$ is around $\sim 0.9$, whereas the coupling of $H_1$, which is
mainly singlet like is $\sim 0.1$. The right-hand side of
Fig.~\ref{fig:cvcoup} shows the relative correction ($x=V,b$)
\beq
\Delta C_x^2 =
\frac{(C_x^2)^{\text{(2loop)}}-(C_x^2)^{\text{(1loop)}}}{(C_x^2)^{\text{(1loop)}}}  
\eeq
when going from one-loop to two-loop. Since the inclusion of the
two-loop corrections changes the admixture of the different Higgs
bosons, the coupling of $H_1$ to vector bosons is reduced, whereas the
coupling of $H_2$ is increased. The relative corrections can be up to
40\%. The couplings $C_u$ to the up-type quarks,
\beq
C_u (H_i) = \frac{{\cal R}^l_{i2}}{\sin\beta} \;,
\eeq
show almost the same behaviour so that we do not display the
corresponding figures separately here. In both cases, the two-loop
corrections render the SM-like Higgs boson even more SM-like. 
\s

Figure~\ref{fig:cbcoup} is the analogous plot for the coupling to the
bottom quarks. As $H_1$ has a non-negligible $h_d$ admixture,
quantified by ${\cal R}^l_{i1}$, for the chosen $\tan\beta\approx 4$ its
coupling to bottom quarks is significant. The 
two-loop corrections reduce this coupling by about 10-20\%. The
$h_u$-like $H_2$ couples with comparable strength to the
down-type quarks at one-loop level, but at two-loop level the
corrections increase the coupling by 
up to 30\%. Figures~\ref{fig:cvcoup} and \ref{fig:cbcoup} illustrate
that the influence of the two-loop corrections on the couplings of the
light Higgs bosons can be sizeable, which in turn leads to significant
effects on the phenomenology of these Higgs bosons. This underlines
the importance of including the two-loop corrections to the Higgs
boson masses and mixing matrix elements for proper phenomenological
investigations. 

\section{Conclusions \label{sec:conclusion}}
We have computed the two-loop corrections to the masses of the Higgs
bosons in the CP-violating NMSSM at order ${\cal
  O}(\al_t \al_s)$ using the Feynman diagrammatic approach with
vanishing external momentum. 
The calculation is based on a mixed $\DRb$-on-shell
renormalization scheme. The corrections have been implemented in the Fortran
package {\tt NMSSMCALC}. The user has the choice between the default
$\DRb$ and an OS scheme for the renormalization of the top/stop
sector. For the light Higgs boson masses, the corrections turn out to
be important and are of the order of 5-10\% for the SM-like Higgs
boson, depending on the adopted top/stop renormalization scheme. The
effect on its couplings to the vector bosons and to the top quarks is
of the same order, with even larger corrections for the smaller bottom
Yukawa couplings. To summarize, the two-loop corrections mainly affect
the mass and the couplings of the $h_u$-dominated Higgs boson as well
as the couplings of the light singlet-like Higgs state. 
For a proper interpretation of the experimental
results and in order to make reliable theoretical predictions, two-loop
corrections therefore have to be taken into account, in particular
when investigating the phenomenology of the light Higgs bosons. The
genuine NMSSM contributions at two-loop order turn out to be small for
 values of the singlet-doublet mixing coupling $\lambda$, that are
still within the perturbativity limit. \s

An estimate of the remaining theoretical uncertainties due to missing
higher order corrections, based on the variation of the
renormalization scheme in the top/stop sector, shows, that the
uncertainty is reduced when going from one- to two-loop order. The
difference in the mass values of the SM-like Higgs boson for the two
schemes decreases from 15-25\% to below 1.5\%. \s

We have not considered yet the ${\cal O} (\al_b\al_s)$ contribution in the
two-loop corrections. It is small for small values of $\tan\beta$, as
chosen here and as favoured by the NMSSM. We  plan to include the 
${\cal O}(\al_b\al_s)$ correction in future work. 

\subsubsection*{Acknowledgments} 
DTN (in part), MMM and KW are supported by the DFG SFB/TR9  ``Computational
Particle Physics''.  DTN thanks the Institute for Theoretical Physics
at the Karlsruhe Institute of Technology for hospitality where part of this
work has been performed. We are grateful to Pietro Slavich and Dominik
St\"ockinger for discussions.

\section*{Appendix}
\begin{appendix}
\section{The running $\DRb$ top mass \label{app:mtoprun}}
Using as input the top quark pole mass $M_t$, we first translate it to
the running $\MSb$ top mass $m^{\MSb}_t(M_t)$ by applying the two-loop
relation, see \eg \cite{Melnikov:2000qh} and references therein, 
\be 
m^{\MSb}_t(M_t)=\left( 1-\fr43\braket{\fr{\al_s(M_t)}{\pi}} -
9.1253\braket{\fr{\al_s(M_t)}{\pi}}^2 \right) M_t \; ,
\ee
where $\alpha_s$ is the strong coupling constant  at two-loop
order. Then $m^{\MSb}_t(M_t)$ is evolved up to the renormalization
scale $\mu_R$, by using the two-loop formula 
\be
m_t^{\MSb}(\mu_R)=U_6(\mu_R, M_t)m^{\MSb}_t(M_t)  \hs \;
\mbox{for} \; \mu_R > M_t  
\; ,
\label{mb_evolution}
\ee   
where the evolution factor $U_n$ reads (see {\it e.g.}~\cite{Carena:1999py})
\beq
U_n(Q_2,Q_1)&=&\left(\fr{\alpha_s(Q_2)}{\alpha_s(Q_1)}\right)^{d_n}\left[1 + 
\fr{\alpha_s(Q_1) - \alpha_s(Q_2)}{4\pi}J_n\right] \; ,\hs Q_2 > Q_1\\ \non
d_n&=&\fr{12}{33-2n} \; , \hs J_n = -\fr{8982 - 504n + 40n^2}{3(33 -
  2n)^2} \; ,
\eeq
with $n=6$ for $Q> M_t$. From the $\MSb$ masses the $\DRb$ masses are 
computed at the SUSY scale, {\it i.e.}~$\mu_R= M_{\text{SUSY}}$, by
using the two-loop relation
\cite{Avdeev:1997sz,Harlander:2006rj,Harlander:2007wh}\footnote{The
  relation is applied at the SUSY scale, where the full supersymmetric
  theory holds and the evanescent coupling $\alpha_e$ can be identified with the
$\overline{\text{DR}}$ coupling $\alpha_s^{\overline{\text{DR}}}$
\cite{Harlander:2006rj,Harlander:2007wh}. The $\overline{\text{DR}}$ coupling
$\alpha_s^{\overline{\text{DR}}}$ is then translated to
$\alpha_s^{\overline{\text{MS}}} \equiv \alpha_s$.},
\beq
m_{t}^{\DRb, \text{SM}}(M_{\text{SUSY}})=m_{t}^{\MSb}(M_{\text{SUSY}})\left[1 -
  \fr{\alpha_s (M_{\text{SUSY}})}{3\pi} -
  \fr{\alpha_s^2 (M_{\text{SUSY}})}{144\pi^2}(73-3n)
\right] \; .\label{eq:SMDRBmt}
\eeq  
The $\DRb$ supersymmetric top mass is then calculated from the $\DRb$
SM top mass as, 
\be
m_t^{\DRb,\text{NMSSM}} = m_t^{\DRb, \text{SM}}(M_{\text{SUSY}}) + dm_t\,,
\ee
where
\begin{align}
dm_t =& \fr{\alpha_s(M_{\text{SUSY}})}{6\pi}\Big[-2m_t\text{Re}\Big(B_1(m_t^2,m_{\tilde{g}}^2,m_{\tilde t_1}^2)+
B_1(m_t^2,m_{\tilde{g}}^2,m_{\tilde t_2}^2) \label{eq:susyqcd_dmt}
\\ \non 
&+  2m_{\tilde{g}}\text{Re}\left(B_0(m_t^2,m_{\tilde{g}}^2,m_{\tilde t_1}^2) -B_0(m_t^2,m_{\tilde{g}}^2,m_{\tilde t_2}^2 ) \right)
\\ \non
&\times(e^{i(\varphi_3+ \varphi_u)} \mathcal{U}_{\tilde t_{22}}
 \mathcal{U}_{\tilde t_{21}}^*+e^{-i(\varphi_3+ \varphi_u)} \mathcal{U}_{\tilde t_{21}} \mathcal{U}_{\tilde t_{22}}^*) \Big]\,.
\end{align}
Here the $\DRb$ top mass at the SUSY-scale has to be used, \ie $m_{t} 
=m_{t}^{\DRb}(M_{\text{SUSY}})$. For the scalar two-point function 
$B_0(p^2,m_1^2,m_2^2)$ we use the convention
\begin{equation}
  B_0(p^2,m_1^2,m_2^2)=16\pi^2\mu_R^{4-D}\int 
\frac{d^Dq}{i(2\pi)^D}\frac{1}{(q^2-m_1^2)((q-p)^2-m_2^2)} \; .
\end{equation}
The two-point tensor integral of rank one $B_1(p^2,m_1^2,m_2^2)$, can be 
written in terms of scalar one-point and two-point functions as
\begin{equation}
  B_1(p^2,m_1^2,m_2^2)=\frac{1}{2p^2}\Big[A_0(m_1^2)-A_0(m_2^2)-
  (p^2-m_2^2+m_1^2) B_0(p^2,m_1^2,m_2^2)\Big] \; ,
\end{equation}
where the convention for $A_0$ is
\begin{equation}
  A_0(m^2)=16\pi^2\mu_R^{4-D}\int
  \frac{d^Dq}{i(2\pi)^D}\frac{1}{(q^2-m^2)} \; .
\end{equation}

\section{Counterterm Mass Matrix}
\label{sec:dMH2loop}
The $\DRb$ counterterms for $|\lambda|$ and $\tan\beta$ 
and furthermore the divergent parts of the OS counter\-terms $\deltatwo v$ 
and $\delta M_{H^\pm}^2$
are related to the counterterm of the field renormalization constant 
$\deltatwo Z_{H_u}$ as already
explained in Sec.~\ref{sec:fixcount}.
If these relations are inserted explicitly into the renormalized 
self-energy, it can be shown analytically
that most of the $\deltatwo Z_{H_u}$ contributions from the counter\-term 
mass matrix cancel against
the field renormalization part of the renormalized self-energy and only 
one additional contribution in the
$h_dh_s$ component is left. Hence, we give the explicit analytic form 
only for the
  part of the counterterm mass matrix
of the neutral Higgs bosons at two-loop level that yields finite
contributions, $\deltatwo\MH\big|^{\text{fin}}$, {\it
   i.e.} counterterms of $\drbar$ parameters are dropped.
Hence $\deltatwo\MH\big|^{\text{fin}}$ only depends on the two-loop
counterterms $\deltatwo M_{H_\pm}^2$, $\deltatwo v$, $\deltatwo
t_{h_u}$, $\deltatwo t_{h_d}$, $\deltatwo t_{h_s}$, $\deltatwo
t_{a_d}$ and $\deltatwo t_{a_s}$ as defined in
Sec.~\ref{sec:fixcount}.
The counterterm mass matrix is given in the basis $(h_d,h_u,h_s,a,a_s)$.
%
%
%
\begin{align}
 \deltatwo\MH\big|^{\text{fin}}_{h_dh_d}=&\deltatwo v\, v |\lambda |^2
 \sin^2\!\beta+\deltatwo M_{H^\pm}^2 \sin^2\!\beta \nonumber\displaybreak[0]\\[2mm]
\quad&+\frac{\deltatwo t_{h_d}
   \left(1-\sin^4\!\beta\right)}{v \cos\!\beta}-\frac{\deltatwo t_{h_u} \sin
   \!\beta \cos^2\!\beta}{v}\displaybreak[0]\,,\\[5mm]
\deltatwo\MH\big|^{\text{fin}}_{h_dh_u}=&\deltatwo v\, v |\lambda |^2 \sin \!\beta \cos \!\beta-\deltatwo M_{H^\pm}^2 \sin \!\beta \cos
   \!\beta+\frac{\deltatwo t_{h_d} \sin^3\!\beta}{v}+\frac{\deltatwo t_{h_u} \cos^3\!\beta}{v}\displaybreak[0]\,,\\[5mm]
\deltatwo\MH\big|^{\text{fin}}_{h_dh_s}=&\frac{\deltatwo v\,
  \left(|\lambda |^2 \cos \!\beta \left(2 v_s^2-3 v^2
      \sin^2\!\beta\right)-\sin \!\beta \left(v_s^2 |\kappa | |\lambda
      | \cos \varphi_y+\sin \!2 \beta
      \,M_{H^\pm}^2\right)\right)}{2v_s} \nonumber\displaybreak[0]\\[2mm]
&\quad-\frac{\deltatwo M_{H^\pm}^2 v \sin^2\!\beta \cos \!\beta
   }{v_s}+\frac{\deltatwo t_{h_d} \sin^4\!\beta}{v_s}+\frac{\deltatwo t_{h_u} \sin
   \!\beta \cos^3\!\beta}{v_s}\displaybreak[0]\,,\\[5mm]
\deltatwo\MH\big|^{\text{fin}}_{h_da\,}=&\frac{\deltatwo t_{a_d} }{v \tan \!\beta}\displaybreak[0]\,,\\[5mm]
\deltatwo\MH\big|^{\text{fin}}_{h_da_s}=&\frac{\deltatwo t_{a_d}}{v_s}-\frac{3}{2} \deltatwo v\, v_s |\kappa | |\lambda | \sin
   \!\beta \sin \varphi_y\displaybreak[0]\,,\\[5mm]
\deltatwo\MH\big|^{\text{fin}}_{h_uh_u}=&\deltatwo v\, v |\lambda |^2 \cos^2\!\beta+\deltatwo M_{H^\pm}^2 \cos^2\!\beta\nonumber\displaybreak[0]\\[2mm]
&\quad-\frac{\deltatwo t_{h_d}
   \sin^2\!\beta \cos \!\beta}{v}+\frac{\deltatwo t_{h_u} (5 \sin \!\beta+\sin 3 \beta
   )}{4 v}\displaybreak[0]\,,\\[5mm]
\deltatwo\MH\big|^{\text{fin}}_{h_uh_s}=&\frac{\deltatwo v\, \left(\sin \!\beta \left(|\lambda |^2 \left(2 v_s^2-3 v^2 \cos^2\!\beta\right)-2 \cos^2\!\beta M_{H^\pm}^2\right)-v_s^2 |\kappa | |\lambda |
   \cos \!\beta \cos \varphi_y\right)}{2
   v_s}\nonumber\displaybreak[0]\\[2mm]
&-\frac{\deltatwo M_{H^\pm}^2 v \sin \!\beta \cos^2\!\beta
   }{v_s}+\frac{\deltatwo t_{h_d} \sin^3\!\beta \cos \!\beta
   }{v_s}+\frac{\deltatwo t_{h_u} \cos^4\!\beta}{v_s}\displaybreak[0]\,,\\[5mm]
\deltatwo\MH\big|^{\text{fin}}_{h_ua\,}=&\frac{\deltatwo t_{a_d}}{v}\displaybreak[0]\,,\\[5mm]
\deltatwo\MH\big|^{\text{fin}}_{h_ua_s}=&\frac{\deltatwo t_{a_d}}{v_s \tan\!\beta}-\frac{3}{2} \deltatwo v\, v_s |\kappa |
   |\lambda | \cos \!\beta \sin \varphi_y\displaybreak[0]\,,\\[5mm]
\deltatwo\MH\big|^{\text{fin}}_{h_sh_s}=&\frac{\deltatwo v v }{2
   v_s^2}\Big(\sin^2\!2 \beta  \big(v^2 |\lambda
   |^2+M_{H^\pm}^2\big)+v_s^2 |\kappa | |\lambda | \big(3 \sin \!2 \beta  \sin
   \varphi_y \tan\! \left(\varphi _{\kappa }+3 \varphi _s\right)\nonumber\displaybreak[0]\\[2mm] 
&-2 \sin
   \!\beta  \cos \!\beta  \cos \varphi_y\big)\Big)+\frac{\deltatwo M_{H^\pm}^2 v^2 \sin ^2\!\beta  \cos ^2\!\beta
   }{v_s^2}\nonumber\displaybreak[0]\\[2mm]
&-\frac{\deltatwo t_{a_d} v \cos \!\beta  \tan\! \left(\varphi _{\kappa }+3 \varphi_s\right)}{v_s^2}+\frac{\deltatwo t_{a_s} \tan \!\left(\varphi _{\kappa }+3 \varphi_s\right)}{v_s}\nonumber\displaybreak[0]\\[2mm]
&-\frac{\deltatwo t_{h_d} v \sin ^4\!\beta  \cos \!\beta
   }{v_s^2}+\frac{\deltatwo t_{h_s}}{v_s}-\frac{\deltatwo t_{h_u} v \sin \!\beta  \cos
   ^4\!\beta }{v_s^2}\displaybreak[0]\,,\\[5mm]
\deltatwo\MH\big|^{\text{fin}}_{h_sa}=&\frac{1}{2} \deltatwo v\, v_s |\kappa | |\lambda | \sin \varphi_y+\frac{\deltatwo t_{a_d} }{v_s\sin\!\beta}\displaybreak[0]\,,\\[5mm]
\deltatwo\MH\big|^{\text{fin}}_{h_sa_s}=&2\deltatwo v\, v |\kappa | |\lambda | \sin \!2 \beta  \sin \varphi_y-\frac{2 \deltatwo t_{a_d} v \cos \!\beta}{v_s^2}+\frac{2
   \deltatwo t_{a_s}}{v_s}\displaybreak[0]\,,\\[5mm]
\deltatwo\MH\big|^{\text{fin}}_{a\,a\,}=&\deltatwo v\, v |\lambda |^2+\deltatwo M_{H^\pm}^2\displaybreak[0]\,,\\[5mm]
\deltatwo\MH\big|^{\text{fin}}_{a\,a_s}=&\frac{\deltatwo v\, \left(\sin \!2 \beta  \left(3 v^2 |\lambda |^2+2M_{H^\pm}^2\right)-6
   v_s^2 |\kappa | |\lambda | \cos \varphi_y\right)}{4
   v_s}+\frac{\deltatwo M_{H^\pm}^2 v \sin \!\beta \cos \!\beta
   }{v_s}\nonumber\displaybreak[0]\\[2mm]
&-\frac{\deltatwo t_{h_d} \sin^3\!\beta}{v_s}-\frac{\deltatwo t_{h_u} \cos
   ^3\!\beta}{v_s}\displaybreak[0]\,,\\[5mm]
\deltatwo\MH\big|^{\text{fin}}_{a_sa_s}=&\frac{\deltatwo v v \sin \!2 \beta  }{2 v_s^2}\Big(\sin \!2 \beta  \big(v^2 |\lambda|^2+M_{H^\pm}^2\big)\nonumber\displaybreak[0]\\[2mm]
&+3 v_s^2 |\kappa | |\lambda | \big(\cos \varphi_y-3 \sin \varphi_y
\tan \left(\varphi_{\kappa }+3 \varphi
  _s\right)\big)\Big)+\frac{\deltatwo M_{H^\pm}^2 v^2 \sin ^2\!\beta  
   \cos ^2\!\beta }{v_s^2}\nonumber\displaybreak[0]\\[2mm]
&+\frac{3 \deltatwo t_{a_d} v \cos \!\beta  \tan
  \!\left(\varphi_{\kappa }+3 \varphi _s\right)}{v_s^2}-\frac{3
  \deltatwo t_{a_s} \tan\! \left(\varphi_{\kappa }+3 \varphi
    _s\right)}{v_s}\nonumber\displaybreak[0]\\[2mm] 
&-\frac{\deltatwo t_{h_d} v \sin ^4\!\beta  \cos
   \!\beta }{v_s^2}+\frac{\deltatwo t_{h_s}}{v_s}-\frac{\deltatwo t_{h_u} v \sin \!\beta
    \cos ^4\!\beta }{v_s^2}\,.
\end{align}

\end{appendix}

\vspace*{1cm}
\bibliographystyle{h-physrev}

\end{document}